\documentclass[aps, prb,superscriptaddress, twocolumn,preprintnumbers,amsmath,amssymb]{revtex4-1}
\usepackage{graphicx}
\usepackage{bm}
\usepackage{amssymb, latexsym, amsmath, textcomp}
\usepackage{verbatim}

\def\bx{\boldsymbol{x}}
\def\by{\boldsymbol{y}}

\def\br{\boldsymbol{r}}
\def\bR{\boldsymbol{R}}

\def\rSU2{{\rm SU}(2)}
\def\prsum{\sum_{(\br, \br')} \, \! \! \! '}

\def\rfs#1{Eq.~(\ref{#1})}

\begin{document}
\title{Topological liquids and valence cluster states in two-dimensional ${\rm SU}(N)$ magnets}
\author{Michael Hermele}
\author{Victor Gurarie}
\affiliation{Department of Physics, University of Colorado, Boulder, Colorado 80309, USA}

\date{\today}
\begin{abstract}
We study the zero temperature phase diagram of a class of two-dimensional SU$(N)$ antiferromagnets.  These models are characterized by having the same type of SU$(N)$ spin placed at each site of the lattice, and share the property that, in general, more than two spins must be combined to form a singlet.  An important motivation to study these systems is that they may be realized naturally in Mott insulators of alkaline earth atoms placed on optical lattices; indeed, such Mott insulators have already been obtained experimentally, although the temperatures are still high compared to the magnetic exchange energy.  We study these antiferromagnets in a large-$N$ limit, finding a variety of ground states.  
Some of the models studied here have a valence bond solid ground state, as was found in prior studies, yet we find that many others have a richer variety of ground states.
Focusing on the two-dimensional square lattice, in addition to valence cluster states (which are analogous to valence bond states), we find both Abelian and non-Abelian chiral spin liquid ground states, which are magnetic counterparts of the fractional quantum Hall effect.  We also find a ``doubled'' chiral spin liquid ground state that preserves time reversal symmetry.  These results are based on a combination of rigorous lower bounds on the large-$N$ ground state energy, and a systematic numerical ground state search.
We conclude by discussing whether experimentally relevant SU$(N)$ antiferromagnets -- away from the large-$N$ limit -- may be  chiral spin liquids.
\end{abstract}

\maketitle

\section{Introduction}
\label{sec:intro}

Since their discovery nearly thirty years ago, fractional quantum Hall (FQH) liquids continue to be a rich source of novel and exciting physics.\cite{tsui82, laughlin83}  FQH liquids belong to an intriguing class of quantum states of matter:  they do not fall into the conventional classification in terms of broken symmetry, electron band structure, and Fermi liquid theory, but instead are characterized by the notion of topological order.  \cite{wen90} $^{\!\!\! \text{,} }$ \footnote{Topological band insulators are characterized by topological properties of electron band structure, which is distinct from topological order as the term is used here.}
In FQH liquids, topological order is directly responsible for the celebrated properties of fractional charge, fractional and non-Abelian statistics, and gapless chiral edge states. While the phenomenon of topological order is not limited to FQH systems in principle, they remain its only known experimental realization, recent progress with
rotating cold atomic condensates notwithstanding.\cite{Gemelke10}  It is thus important to ask in which other systems we might find topologically ordered states of matter.

Over the past several years there has been considerable progress identifying model quantum spin systems exhibiting topological order.\cite{moessner01, balents02, senthil02, motrunich02, kitaev02, freedman04, levin05, kitaev06}  The ground states of these models typically have no spontaneously broken symmetries, and are thus  referred to as quantum spin liquids; these are concrete realizations of Anderson's idea of resonating valence bonds.\cite{anderson73}  The models that can be shown to exhibit topological order are generally not very realistic, but many are built from realistic degrees of freedom without any special symmetries, making it clear that there is no in-principle obstacle for topological order to exist in real quantum spin systems.  Despite this progress, we know of no candidate materials for a topologically ordered spin liquid.  While a few solid state quantum magnets are quantum spin liquid candidates,\cite{shimizu03, helton07, ofer06, mendels07, okamoto07, itou08} all these systems seem to have gapless excitations and are thus not natural candidates for topological order.

Here, we discuss a class of spin systems with topologically ordered spin liquid ground states.  While the systems we study are not realistic for solid state materials, they can be realized naturally -- without complicated engineering of a special Hamiltonian -- using fermionic ultracold alkaline earth atoms (AEA) in optical lattices.\cite{gorshkov10}  While most ultracold atom experiments to date involve alkali atoms, AEA are promising systems to study many-body physics, and experiments in this direction are progressing rapidly.\cite{fukuhara07a, fukuhara07b, fukuhara09, escobar09, stellmer09, desalvo10, taie10, tey10, stellmer10, sugawa10}  An important feature of these systems is the presence -- without fine-tuning -- of a large ${\rm SU}(N)$ spin-rotation symmetry, where $N = 2I + 1$, and $I$ is the nuclear spin.\cite{gorshkov10, cazalilla09}  The nuclear spin can be as large as $I = 9/2$ for  $^{87}$Sr, so $N$ can be as large as 10.  The focus of this paper is primarily on the models themselves, and not their cold atom realizations; nonetheless, for completeness we review in Appendix~\ref{sec:atomic} the realization of the spin systems of interest using AEA.  Further information along these lines can be found in Ref.~\onlinecite{gorshkov10}.

\begin{figure}
\includegraphics[width=1.5in]{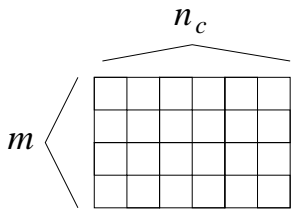}
\caption{The Young tableau corresponding to the $m \times n_c$ irreducible representation of ${\rm SU}(N)$.  We consider spin models where the spin at each lattice site transforms in this representation.}
\label{fig:rect_youngtab}
\end{figure}

The models we study are two-dimensional ${\rm SU}(N)$ antiferromagnets where the ${\rm SU}(N)$ representation (\emph{i.e.} type of spin) is the same on every lattice site -- the simplest case is the $N$-dimensional ${\rm SU}(N)$ fundamental representation. More generally, we consider spins in the ${\rm SU}(N)$ irreducible representation labeled by a $m \times n_c$ Young tableau with $m < N$ rows and $n_c$ columns (Fig.~\ref{fig:rect_youngtab}).  We will refer to such a representation as the $m \times n_c$ representation.
These models differ crucially from solid state ${\rm SU}(2)$ magnetism, in that, in general, more than two spins are required to form a ${\rm SU}(N)$ singlet.  This means that singlets are not two-site valence bonds, but rather are multi-site ``valence clusters.''  The cases $n_c = 1$ and $n_c = 2$ can both be realized as AEA Mott insulators (Appendix~\ref{sec:atomic}).  While a variety of values of $m$ can be realized, $m=1$ is of greatest interest because it best avoids issues of three-body and other losses.  For ${\rm SU}(2)$ spins, $m = 1$ and $n_c = 2S$, so that $n_c = 1,2$ correspond to $S = 1/2, 1$, respectively.  It should be noted that these models are distinct from a much-studied class of ${\rm SU}(N)$ spin models, where (inequivalent) conjugate representations occupy the two sublattices of a bipartite lattice.\cite{read89a, read89b}

In Ref.~\onlinecite{hermele09}, together with A. M. Rey, we considered the semiclassical limit $m=1$ and $n_c \to \infty$.  This is analogous to the large-$S$ limit for ${\rm SU}(2)$ spins, and, indeed, reduces to it when $N = 2$.  The limit $n_c \to \infty$ is biased toward magnetically ordered ground states, because the spins become \emph{classical} $N$-component complex  vectors.  However, it turns out that the ground state manifold is in general extensively degenerate (precisely, its dimension is proportional to the number of sites in the system) -- on the square lattice with nearest-neighbor exchange this occurs for $N \geq 3$, with the degree of extensive degeneracy growing with $N$.  This situation sometimes occurs in geometrically frustrated magnets, where a common consequence is that magnetic order is strongly suppressed, and sometimes even destroyed, by large very low-energy fluctuations.\cite{moessner98}  Given that this occurs even in a limit which is deliberately biased in favor of magnetic order, non-magnetic ground states are likely for the $n_c = 1,2$ cases of greatest interest.  (We note that recent work has given strong evidence that a magnetically ordered ground state does occur for $m = n_c = 1$ and $N = 3$ on the square lattice.\cite{toth10}  For $N=4$, while prior exact diagonalization\cite{bossche00} and variational wavefunction\cite{wangf09} studies favored a non-magnetic ground state, a very recent study employing both projected entangled pair states and exact diagonalization found evidence in favor of magnetic order.\cite{corboz11} These results are consistent with the expectation that non-magnetic ground states are more likely for larger values of $N$, where the extensive degeneracy in the semiclassical limit is larger.)

This paper is concerned with the ground states of these ${\rm SU}(N)$ antiferromagnets, in a solvable large-$N$ limit suitable for addressing the competition among non-magnetic states.\cite{affleck88, marston89}  With Rey in Ref.~\onlinecite{hermele09}, we studied the case $n_c = 1$ on the square lattice in the large-$N$ limit, and announced a number of results.  Here, we study the case of arbitrary $n_c$ on general lattices, with a focus on the square lattice for $n_c = 1,2$.  We also provide more detail on the results already reported in Ref.~\onlinecite{hermele09}.

In the large-$N$ limit, $N$ and $m$ are taken to infinity, while $N/m = k$ and $n_c$ are held fixed.  The parameter $k$, which we choose to be an integer greater than unity, plays a very important role in our analysis:  $k$ \emph{is the minimum number of spins needed to form a} ${\rm SU}(N)$ \emph{singlet}.   Given this physical interpretation of $k$, the large-$N$ limit can thought of as a solvable generalization of the model with ${\rm SU}(k)$ symmetry, $m=1$, and the same fixed $n_c$.  Readers primarily interested in the implications of our results for real AEA Mott insulators can interpret our large-$N$ results as a prediction for the ground state of these physically realizable $m = 1$ models.  This bold prediction will need to be tested further in future work; see Sec.~\ref{sec:discussion} for further discussion along these lines.

In general, the ${\rm SU}(N)$ singlets are $k$-site valence clusters.  Based on the observation that, when $k = 2$, singlets are 2-site valence bonds, the case $k = 2$ has been  studied as a solvable large-$N$ generalization of ${\rm SU}(2)$ antiferromagnetism.\cite{affleck88, marston89, read89b, rokhsar90}  For the same reason, the $k > 2$ case does not provide a good generalization of ${\rm SU}(2)$ antiferromagnetism.  Under very general conditions in the $k=2$ large-$N$ limit, the ground state is a  valence-bond solid (VBS)  that spontaneously breaks lattice symmetries.\cite{rokhsar90}  One of the striking results of this paper (and Ref.~\onlinecite{hermele09}) is that the large-$N$ ground states are much richer in the less-studied case $k > 2$.  

While ${\rm SU}(N)$ spin models with the same representation on every lattice site have not received extensive attention (except in the case of self-conjugate representations, \emph{i.e.} $k = 2$), there have been several earlier studies.  While our focus is primarily on two dimensions, we note that the one-dimensional chain with $m = n_c = 1$ was solved exactly for all $N$,\cite{sutherland75} and the effective field theories of it and other chains were also studied.\cite{affleck88a} In two dimensions, most work focused on the $m = n_c = 1$ model with either $N = 3$ or $N = 4$.  The former case arises as a special point of a $S = 1$ spin model with bilinear and biquadratic exchange terms,\cite{lauchli06,toth10} while the latter is a highly symmetric point of a $S = 1/2$ Mott insulator with an additional two-fold orbital degeneracy,\cite{pokrovskii72,li98, bossche00, wangf09} or a special point of a model with ${\rm Sp}(4)$ symmetry.\cite{wucj03}  We also note a further very recent study of the $N=4$, $m = n_c = 1$ model on the square lattice.\cite{corboz11} Models on the cubic lattice have been studied in high-temperature series expansion,\cite{fukushima05} and a class of exactly solvable models with $n_c > 1$ was studied in Ref.~\onlinecite{arovas08}.  Finally, effective models of valence cluster degrees of freedom -- analogous to more familiar quantum dimer models -- have been studied.\cite{pankov07, xuc08}
  
Returning for a moment to the ultracold atom realization of our models, it should be mentioned that high-spin quantum magnets can also be realized using alkali atoms, and in that context also have spin symmetry enhanced above ${\rm SU}(2)$.\cite{wucj03}  However for an $N$-component system, the symmetry is generically less than ${\rm SU}(N)$.  For example, in a spin-$3/2$ alkali system, the spin symmetry is expected generically to be ${\rm Sp}(4)$ and not ${\rm SU}(4)$.\cite{wucj03}  While these systems share with ${\rm SU}(2)$ magnets the property that two spins can be combined to form a singlet, they are also likely to be fertile ground for the realization of a variety of interesting ground states.\cite{wucj03,chen05,lecheminant05, wucj05, wucj06, szirmai10, wucj10} 
In close relation to quantum magnetism, half-filled repulsive ${\rm SU}(N)$ Hubbard models have also been studied in the context of ultracold atoms.\cite{honerkamp03, xuc10}  Very recently, the repulsive ${\rm SU}(3)$ Hubbard model was studied for arbitrary filling.\cite{rapp11}

We now summarize our results for the square lattice with $n_c = 1,2$.  (A graphical summary for $n_c = 1$ can be found in the phase diagram of Fig.~\ref{fig:phasediagram}, discussed in Sec.~\ref{sec:discussion}.)  Depending on $k$, we find valence cluster states (VCS) that break lattice symmetries and are formed by tiling the lattice with multi-site singlet clusters, and three distinct types of topologically ordered spin liquids. Two of the spin liquids are chiral spin liquid (CSL) states.\cite{kalmeyer87, kalmeyer89, wen89}  The CSL is a spin system analog of an FQH state;  it spontaneously breaks parity and time-reversal symmetry (symmetries that are broken explicitly by the magnetic field in FQH systems),\footnote{The CSL is still referred to as a spin liquid, even though it spontaneously breaks parity and time-reversal.} supports excitations with fractional quantum numbers and statistics, and has gapless chiral edge states that carry spin.  

  We find both an Abelian chiral spin liquid (ACSL) (for $n_c = 1$ and $k \geq 5$), and a non-Abelian Chiral spin liquid (nACSL) (for $n_c = 2$ and $k \geq 6$).  The ACSL is described at low energies by a ${\rm U}(1)_N$ Chern-Simons theory and has fractional statistics.  The nACSL, on the other hand, is described by a ${\rm U}(1)_{2 N} \times {\rm SU}(2)_N$ Chern-Simons theory, and supports non-Abelian statistics.   The third state we find is distinct from these CSL states in that it preserves time reversal symmetry.  At the mean-field level it appears as two copies of the ACSL, with opposite chiralities, and we thus dub it a doubled chiral spin liquid (dCSL); its low-energy theory is a ${\rm U}(1)$ mutual Chern-Simons theory.  A more concrete way of describing all these states is in terms of Gutzwiller projected trial wavefunctions, as described in Sec.~\ref{sec:top}.
The dCSL is found for $n_c = 2$ and has the same energy as the nACSL in the $N \to \infty$ limit.  Presumably $1/N$ corrections select one of these states as a ground state; we have not computed these, since in our view the large-$N$ limit is primarily useful as a tool to determine \emph{likely} ground states of physically realizable models, and ultimately the issue of whether the dCSL or nACSL (or some other state) is lower in energy will need to be determined by directly studying those models (occurring at finite $N$).  We find VCS states for $n_c = 1,2$ and $2 \leq k \leq 4$, as well as a more complicated inhomogeneous ground state when $n_c = 2$ and $k = 5$.  These results are obtained via a combination of exact lower bounds on the large-$N$ ground state energy (generalizing the results of Ref.~\onlinecite{rokhsar90}), and a systematic numerical search.

A current interest in states which support excitations with non-Abelian statistics is fueled by the expectation that they could be used to build a topologically protected quantum computer. \cite{kitaev02} The simplest non-Abelian statistics is described by an ${\rm SU}(2)_2$ Chern-Simons theory. It is believed to be realized in the quantum Hall effect at the
filling fraction $5/2$,  \cite{moore91,willett10} as well as in a variety of setups involving Majorana fermions. \cite{read00,FuKane3d,Lutchyn10,Oreg10,Sau2DEG,Alicea2DEG,Alicea10}  However, it is not rich enough to support universal quantum computations. \cite{nayak08} Non-Abelian statistics described by ${\rm SU}(2)_N$ Chern-Simons theory for $N>2$ is significantly richer and in fact gets richer as $N$ increases. In particular, $N=3$ or $N \ge 5$ is known to be sufficient for universal quantum computations. \cite{freedman02} Some fractional quantum Hall states in the first excited Landau level are believed to  realize these types of non-Abelian statistics for moderate $N$, at least for $N=3$. \cite{read99,pan08} We observe that the non-Abelian statistics proposed here can be as high as ${\rm SU}(2)_{10}$ (in case of $^{87}$Sr), thus it is inherently very rich.

We note that solvable spin models with CSL ground states have been found previously.\cite{khveshchenko89, khveshchenko90, yao07, schroeter07, greiter09, thomale09, greiter11}  One of these\cite{yao07} is a generalization of the Kitaev model on a decorated honeycomb lattice.  The models of Refs.~\onlinecite{schroeter07, greiter09, thomale09, greiter11} involve long-range 6-spin interactions.  Refs.~\onlinecite{khveshchenko89, khveshchenko90} found that a CSL was the large-$N$ ground state of a ${\rm SU}(N)$ spin model of a different type from those considered here, where in addition a four-spin ring exchange term which explicitly broke  time-reversal invariance was added to the Hamiltonain.  Very recently, also motivated by ultracold alkaline earth atoms, Szirmai \emph{et. al.} studied the same type of spin model discussed in this paper, for $n_c = 1$ and $k = 6$ on the honeycomb lattice, and found a CSL ground state in the large-$N$ limit.\cite{szirmai11}

We now give an outline of our paper.  In Sec.~\ref{sec:models}, we define a broad class of SU$(N)$ Heisenberg models in terms of slave fermions; this is convenient for understanding the large-$N$ limit, which is also described in this section. In Sec.~\ref{sec:top} we discuss the properties of the Abelian chiral spin liquid, non-Abelian chiral spin liquid, and doubled chiral spin liquid, including their wavefunctions and edge states. We spend the rest of the paper arguing that these states indeed appear in the large $N$ limit of the appropriate models. In particular, in Sec.~\ref{sec:genlat} we discuss the solution to the large $N$ limit of our models on general lattices. We give examples of lattices where the large $N$ solution can be proven to be a
VCS, by generalizing results of Ref.~\onlinecite{rokhsar90} to $k > 2$.  The principal tool of analysis is a rigorous lower bound on the large-$N$ ground state energy, which is saturated by certain VCS states.
 We also give general arguments that VCS are not the only states which are possible on generic lattices, and other states, including spin liquid states, should naturally appear in appropriate cases.
 In Sec.~\ref{sec:bipartite} we specialize to bipartite lattices, showing that a stricter lower bound on the energy can be obtained in this case (when $k > 2$).  Finally, in Sec.~\ref{sec:square} we further specialize to the square lattice.  Using the rigorous lower bounds, we show that the large-$N$ ground state is a VCS for $k = 2,3,4$, for both $n_c = 1$ and $n_c = 2$.  Next, employing a numerical analysis we show that the large-$N$ ground state at $n_c=1$ on the square lattice is the ACSL for $5 \le k \le 8$.  Moreover, at $n_c=2$ we show that nACSL and dCSL are the degenerate ground states at $k=6, 7$.   Closely tied to these results is the discussion of Appendix~\ref{app:largek}, where we discuss the possible ground states in the limit of large $k$. In particular, for $n_c = 1$, we show that the ACSL wins over VCS states as well as a trial uniform gapless state, giving us ammunition to conjecture that ACSL is the ground state for all $k \ge 5$.  The same analysis goes over to $n_c = 2$ and leads us to conjecture that the nACSL and dCSL are degenerate ground states for all $k \ge 6$. 
 
We would not have studied these models if it were not for the strong potential to realize them in systems of AEA on optical lattices.   The paper concludes with a discussion in Sec.~\ref{sec:discussion}, focusing on the prospects to find chiral spin liquids in those spin models that can be realized in cold atom experiments.  In particular, we discuss the phase diagram in the $k$-$m$ plane (Fig~\ref{fig:phasediagram}).  Finally, we mention some directions for future study; one such direction is to understand how fractional or non-Abelian particles may be localized and braided in these systems, with an eye toward detection of fractional or non-Abelian statistics.  In Appendix~\ref{app:carriers} we further discuss some ideas in this direction, describing how fractional \emph{holons} (which carry conserved atom number but not spin), may be localized by applying an external potential.

 In Appendix~\ref{sec:atomic}, we review some aspects of the cold atom realizations of these systems.  Moreover, starting from the Hubbard model describing AEA on an optical lattice in the large-$U$ limit, we derive the appropriate Heisenberg models using degenerate perturbation theory.  Some technical details are given in Appendices~\ref{app:1site-irrep}, \ref{app:2site-exact} and~\ref{app:kcluster}.

\section{Models and large-$N$ limit}
\label{sec:models}

Here we introduce the ${\rm SU}(N)$ spin models and construct the solvable large-$N$ limit, which allows us to address the competition among non-magnetic ground states.  We shall define the models in terms of the fermionic spinon operators $f^\dagger_{\br a \alpha}$.  Here $\br$ labels lattice sites, $\alpha = 1,\dots,N$ is the ${\rm SU}(N)$ spin index, and $a = 1,\dots,n_c$ will be called a ``color'' index.  The $\alpha$ index transforms in the fundamental representation of ${\rm SU}(N)$ spin rotations; that is, a  global ${\rm SU}(N)$ rotations acts by
\begin{equation}
f_{\br a \alpha} \to U_{\alpha \beta} f_{\br a \beta} \text{,}
\end{equation}
where $U$ is an arbitrary ${\rm SU}(N)$ matrix.  (Here, and throughout the paper, summation over repeated indices is implied.  This does not apply to repeated site labels $\br$.) Similarly, the $a$ index transforms in the fundamental representation of ${\rm SU}(n_c)$ color rotations.  It is important to distinguish the spinons from the physical fermions of an underlying Hubbard model (as in Appendix~\ref{sec:atomic});  we elaborate on this distinction and its importance below.

Before defining the Hamiltonian, we 
must specify the type of spin at each lattice site.  This is accomplished by a pair of local constraints,
\begin{eqnarray}
f^\dagger_{\br a \alpha} f^{\vphantom\dagger}_{\br a \alpha} &=& n_c m  \label{eqn:num-constraint} \\
f^\dagger_{\br a \alpha} T^A_{a b} f^{\vphantom\dagger}_{\br b \alpha} &=& 0 \label{eqn:color-constraint} \text{.}
\end{eqnarray}
Here, $A = 1,\dots,n_c^2 - 1$ labels the traceless, Hermitian $n_c \times n_c$ matrices $T^A$ that generate infinitesimal ${\rm SU}(n_c)$ rotations.  The proper interpretation of these constraints is that, for each lattice site, we restrict to the subspace of the fermion Hilbert space spanned by eigenstates of the left-hand sides of Eqs.~(\ref{eqn:num-constraint},\ref{eqn:color-constraint}), with eigenvalues given by the right-hand sides.  The first constraint specifies a fixed number of fermions on each lattice site, and the second constraint dictates that each site is a color singlet.  The second constraint is omitted when $n_c = 1$.  These constraints project out the ``charge'' (conserved number) and color degrees of freedom of the spinons, which are not physical at the microscopic level but are important for understanding the low-energy effective theories obtained in the large-$N$ limit.  While the constraint Eq.~(\ref{eqn:color-constraint}) may appear mysterious, in the case $n_c = 2$ it arises naturally in the large-$U$ limit of the Hubbard model describing one type of AEA Mott insulator, as described in Appendix~\ref{sec:atomic}.

Taken together, the constraints imply that the spin at each site transforms in the ${\rm SU}(N)$ irreducible representation with a $m \times n_c$ rectangular Young tableau -- this is shown in Appendix~\ref{app:1site-irrep}.  Since \emph{all} physical operators must commute with these local constraints, which together form a ${\rm U}(n_c)$ algebra, in this choice of variables, there is a local ${\rm U}(n_c)$ redundancy.  This is intimately related to the fact that, in the large-$N$ limit, the low-energy effective theory is a ${\rm U}(n_c)$ gauge theory; we shall see this below.  This type of slave particle representation has been employed before.\cite{read89b, freedman04, xuc10}

The ${\rm SU}(N)$ spin operators are defined to be
\begin{equation}
S_{\alpha \beta}(\br) = \sum_a f^\dagger_{\br a \alpha} f^{\vphantom\dagger}_{\br a \beta} \text{,}
\end{equation}
and the Hamiltonian is 
\begin{equation}
\label{eqn:hspin2}
{\cal H} =  \sum_{( \br, \br' )} J_{\br \br'} S_{\alpha \beta}(\br) S_{\beta \alpha} (\br') \text{.}
\end{equation}
Here, the sum is over all pairs of sites $(\br, \br')$.  The cases $n_c = 1,2$ are realizable with alkaline earth atoms, as discussed in Appendix~\ref{sec:atomic}, and most of our analysis is focused on these cases.  We shall always consider $m = N / k$, where $k \geq 2$ is an integer.  The parameter $k$, as introduced in Sec.~\ref{sec:intro}, is the minimum number of spins required to form a ${\rm SU}(N)$ singlet.  This model becomes exactly solvable in the limit where $N$ and $m$ are taken to infinity, while $k$ and $n_c$ are held fixed.  For technical convenience, we also write $J_{\br \br'} = {\cal J}_{\br \br'} / N$ and hold ${\cal J}_{\br \br'}$ fixed;  this corresponds merely to multiplication of the Hamiltonian by a constant.  The case of greatest experimental interest is $m=1$, and the large-$N$ limit with fixed $k$ and $n_c$ should be thought of as a solvable limit of the model with $m=1$ and $N = k$; the minimum number of spins required to form a singlet is the same as in this model. 

We now describe in detail the large-$N$ solution, which follows the work of Affleck and Marston\cite{affleck88,marston89}, and also Read and Sachdev.\cite{read89b}  Affleck and Marston studied the case $n_c = 1$ and $m = N/2$, while Read and Sachdev generalized their results to arbitrary $n_c$ while still fixing $m = N/2$.  The formal structure of the large-$N$ solution is the same as in the earlier works, but, as is discussed in the following sections, the nature of the ground states is dramatically different.

We first consider separately the case $n_c = 1$ for its greater simplicity.  The starting point is the imaginary-time functional integral for the partition function
\begin{equation}
\label{eqn:pf}
Z = \int {\cal D} f {\cal D} \bar{f} {\cal D} \chi {\cal D} \lambda \exp\big( - S(f, \bar{f}, \chi, \lambda) \big) \text{,}
\end{equation}
where the action is
\begin{eqnarray}  \label{eq:hsa}
S &=& \int_0^\beta d\tau \sum_{\br} \Big[ \bar{f}_{\br \alpha} \partial_{\tau} f_{\br \alpha}
+ i \lambda_{\br} \big( \bar{f}_{\br \alpha}  f_{\br \alpha} - m \big) \Big]\label{eqn:nc1action} \\
&+&  \int_0^\beta d\tau\prsum \frac{N}{{\cal J}_{\br \br'} } | \chi_{\br \br'} |^2 \nonumber \\
&+& \int_0^\beta d\tau \prsum \big( \chi_{\br \br'} \bar{f}_{\br \alpha} f_{\br' \alpha} + \text{H.c.} \big) \nonumber \text{.}
\end{eqnarray}
Here, the fermionic variables $f_{\br \alpha}(\tau)$ and $\bar{f}_{\br \alpha}(\tau)$ are the usual Grassmann variables.  $\lambda_{\br}(\tau)$ is a real Lagrange multiplier field that implements the constraint $f^\dagger_{\br \alpha} f^{\vphantom\dagger}_{\br \alpha} = m$.  The primed sum $\sum'_{(\br, \br')}$ in the last two terms is over only those bonds $(\br, \br')$ where ${\cal J}_{\br \br'} \neq 0$, and  $\chi_{\br \br'}(\tau)$ is a complex field defined on the same set of bonds. Upon integrating out $\chi$ one obtains the Hamiltonian Eq.~(\ref{eqn:hspin2}), which is quartic in fermion operators.  We focus on the zero-temperature limit $\beta \to \infty$.

We can formally integrate out the fermions and obtain an effective action
\begin{eqnarray}
S_{{\rm eff}}(\chi, \lambda) &=& \int_0^\beta d\tau\prsum \frac{N}{{\cal J}_{\br \br'} } | \chi_{\br \br'} |^2 - i m \int_0^\beta d\tau \sum_{\br} \lambda_{\br} \nonumber \\  &+& N \operatorname{Tr} \operatorname{ln} Q(\chi, \lambda) \text{,}
\end{eqnarray}
where $Q$ is the quadratic form characterizing the fermionic part of the action Eq.~(\ref{eqn:nc1action}).  Since $m = N/k$, $S_{{\rm eff}}$ has a prefactor of $N$ and no other $N$-dependence, implying that when $N \to \infty$ the functional integral over $\chi$ and $\lambda$ can be done exactly using the saddle point approximation.  We therefore replace $\chi$ and $\lambda$ by non-fluctuating fields
\begin{eqnarray}
\chi_{\br \br'} &\to& \bar{\chi}_{\br \br'} \\
\lambda_{\br} &\to& i \mu_{\br} \text{,}
\end{eqnarray}
which are substituted into Eq.~(\ref{eqn:nc1action}) to obtain a theory of non-interacting fermions subject to the mean-field Hamiltonian
\begin{eqnarray}
{\cal H}_{{\rm MFT}} &=& \prsum \frac{N}{{\cal J}_{\br \br'} } | \bar{\chi}_{\br \br'} |^2 + m \sum_{\br} \mu_{\br}  \\
&+& \prsum \big( \bar{\chi}_{\br \br'} f^\dagger_{\br \alpha} f^{\vphantom\dagger}_{\br' \alpha} + \text{H.c.} \big) - \sum_{\br} \mu_{\br} f^\dagger_{\br \alpha} f^{\vphantom\dagger}_{\br \alpha} \nonumber \text{.}
\end{eqnarray}
The imaginary saddle point $\lambda_{\br} \to i \mu_{\br}$ is needed for the mean-field Hamiltonian to be Hermitian.  We emphasize that despite the appearance of the term ``mean-field,'' and the appearance of mean-field equations and a mean-field Hamiltonian, we are not making any sort of mean-field \emph{approximation}.  
That is, in the $N \to \infty$ limit, the specific mean-field decoupling we consider -- and \emph{only} this decoupling -- becomes exact.  The results we present are thus exact for the Heisenberg spin model in the $N \to \infty$ limit.

In order for this to be a legitimate saddle point, we must satisfy the extremum condition
\begin{equation}
\frac{\delta}{\delta \chi_{\br \br'}} S_{{\rm eff}} \Big|_{\chi \to \bar{\chi}, \lambda \to i \mu}
= \frac{\delta}{\delta \lambda_{\br}} S_{{\rm eff}} \Big|_{\chi \to \bar{\chi}, \lambda \to i \mu}
= 0 \text{.} \label{eqn:excond}
\end{equation}
In the low-temperature limit, the ground state energy $E_{{\rm MFT}}$ of ${\cal H}_{{\rm MFT}}$ satisfies $S_{{\rm eff}}(\chi \to \bar{\chi}, \lambda \to i \mu) = \beta E_{{\rm MFT}}$, so satisfying Eq.~(\ref{eqn:excond}) is equivalent to extremizing the ground state energy.  The saddle point equations of Eq.~(\ref{eqn:excond}) are equivalent to the more convenient expressions
\begin{eqnarray}
\label{eqn:nc1-spe-chi}
\bar{\chi}_{\br \br'} &=& - \frac{{\cal J}_{\br \br'}}{N} \langle f^{\dagger}_{\br' \alpha} f^{\vphantom\dagger}_{\br \alpha} \rangle \\
\label{eqn:nc1-spe-constraint}
m &=& \langle f^{\dagger}_{\br \alpha} f_{\br \alpha}  \rangle  \text{.}
\end{eqnarray}
Here, the expectation values are calculated using the non-interacting Hamiltonian $H_{{\rm MFT}}$.

Now that we have discussed the simpler case $n_c = 1$, we discuss the general case $n_c > 1$, describing only those aspects that differ from $n_c = 1$, and making some definitions that will be useful later on.  The principal difference is that now the fields $\chi$ and $\lambda$ are $n_c \times n_c$ matrices:  $\chi^{a b}_{\br \br'}(\tau)$ is a general complex $n_c \times n_c$ matrix, and $\lambda^{a b}_{\br}(\tau)$ is a $n_c \times n_c$ Hermitian matrix.  The partition function is of the same form as Eq.~(\ref{eqn:pf}), where the integration over $\lambda$ is understood to be over the restricted space of Hermitian matrices.  
The action is now
\begin{eqnarray}
S &=& \int_0^\beta d\tau \sum_{\br} \Big[  \bar{f}_{\br a \alpha} \partial_{\tau} f_{\br a \alpha}  + i \big( \lambda^{b a}_{\br}  \bar{f}_{\br a \alpha} f_{\br b \alpha} - m \operatorname{tr}(\lambda_{\br}) \big) \Big] \nonumber \\
&+& \int_0^\beta d\tau \prsum \frac{N}{{\cal J}_{\br \br'}} \operatorname{tr} ( \chi^\dagger_{\br \br'} \chi^{\vphantom\dagger}_{\br \br'} ) \label{eqn:nc2action} \\
&+& \int_0^\beta d\tau \prsum \big[  \chi^{a b}_{\br \br'} \bar{f}_{\br a \alpha} f_{\br' b \alpha} + \text{H.c.} \big]  \text{.} \nonumber
\end{eqnarray}
The traces in this expression are in the color space.  The field $\lambda_{\br}$ is again a Lagrange multiplier, now implementing both the constraints of Eq.~(\ref{eqn:num-constraint}) and~(\ref{eqn:color-constraint}).  Again, integrating out $\chi_{\br \br'}$ we obtain the Hamiltonian Eq.~(\ref{eqn:hspin2}).

The saddle point values of the fields take the form
\begin{eqnarray}
\chi^{a b}_{\br \br'} &\to& \bar{\chi}^{a b}_{\br \br'} \\
\lambda^{a b}_{\br} &\to& i \mu^{a b}_{\br} \text{,}
\end{eqnarray}
where $\mu_{\br}$ is a Hermitian matrix.  The mean-field Hamiltonian is
\begin{eqnarray}
{\cal H}_{{\rm MFT}} &=& \prsum \frac{N}{{\cal J}_{\br \br'} } \operatorname{tr} ( \bar{\chi}^\dagger_{\br \br'} \bar{\chi}^{\vphantom\dagger}_{\br \br'} ) + m \sum_{\br} \operatorname{tr}(\mu_{\br})  \label{eqn:hmft}  \\
&+& {\cal H}_K + {\cal H}_V \text{,} \nonumber
\end{eqnarray}
where
\begin{eqnarray}
\label{eqn:hk}
{\cal H}_K &=& \prsum \big( \bar{\chi}^{a b}_{\br \br'} f^\dagger_{\br a \alpha} f^{\vphantom\dagger}_{\br' b \alpha} + \text{H.c.} \big) \\
\label{eqn:hv}
{\cal H}_V &=& - \sum_{\br} \mu^{b a}_{\br} \hat{n}^{a b}_{\br}  \text{.}
\end{eqnarray}
Here, we have defined the color density
\begin{equation}
\hat{n}^{a b}_{\br} = f^\dagger_{\br a \alpha} f^{\vphantom\dagger}_{\br b \alpha} \text{.}
\end{equation}
Note that we can also write ${\cal H}_V = - \sum_{\br} \operatorname{tr} ( \mu_{\br} \hat{n}_{\br} )$.
The saddle point equations are now
\begin{eqnarray}
\bar{\chi}^{a b}_{\br \br'} &=& - \frac{{\cal J}_{\br \br'}}{N} \langle f^{\dagger}_{\br' b \alpha} f^{\vphantom\dagger}_{\br a \alpha} \rangle \label{eqn:spe-chi} \\
m \, \delta^{a b} &=& \langle \hat{n}^{a b}_{\br} \rangle \label{eqn:spe-constraints} \text{.}
\end{eqnarray}
When analyzing the mean-field Hamiltonian we shall always work in the canonical ensemble for the conserved fermion number.

This discussion shows that finding the ground state in the large-$N$ limit reduces to finding the saddle point with lowest energy $E_{{\rm MFT}}$.  In general this task, while a great deal simpler than finding the ground state of the original quantum problem, is still nontrivial.  We shall make progress below using a combination of exact lower bounds on $E_{{\rm MFT}}$, and numerical methods to search for ground states.

As mentioned above, it is important to recognize that the $f^\dagger_{\br a \alpha}$ spinon operators are not the same as the physical fermions of the Hubbard model discussed in Appendix~\ref{sec:atomic}.  We are describing a spin model, and there are only ${\rm SU}(N)$ spin degrees of freedom.  In addition to spin degrees of freedom the physical alkaline earth atom fermions have degrees of freedom associated with their conserved number, as well as with their $^{1}S_0$ and $^{3}P_0$ electronic states.  These degrees of freedom are not present in the model; this is appropriate for a low-energy description of the Mott insulating states we are describing, where excitations associated with these degrees of freedom are gapped.  (To describe such gapped excitations, one must return to the original Hubbard model.)

The difference between the spinons and the physical fermions is manifest when we consider the fluctuations about a mean-field saddle point.  This allows us to construct a low-energy effective theory, which goes beyond mean-field theory for a given saddle point.  The spinons in this effective theory should not be interpreted as a microscopic representation of the spins, but as low-energy effective degrees of freedom.  As we shall see below, the spinons are minimally coupled to a fluctuating ${\rm U}(n_c)$ gauge field.  On the other hand, the physical fermions of the underlying Hubbard model are uncharged under this ${\rm U}(n_c)$ gauge field and do not couple to it directly.

We consider fluctuations of the form
\begin{eqnarray}
\lambda^{a b}_{\br}(\tau) &=& i \mu^{a b}_{\br} + a^{a b}_{\br \tau}(\tau) \\
\chi^{a b}_{\br \br'}(\tau) &=& \Big[ \bar{\chi}_{\br \br'} \exp \big( i a_{\br \br'}(\tau) \big) \Big]^{a b} \text{,}
\end{eqnarray}
where $a_{\br \br'}(\tau)$ is a $n_c \times n_c$ Hermitian matrix, so that $e^{i a_{\br \br'} }$ is unitary.  While other fluctuations are typically trivially massive (\emph{e.g.} amplitude fluctuations in $\chi$), these fluctuations take the form of a ${\rm U}(n_c)$ gauge field minimally coupled to the spinons.  Specifically, $a_{\br \tau}$ and $a_{\br \br'}$ form the time and space components, respectively, of the fluctuating ${\rm U}(n_c)$ vector potential.  Gauge fluctuations can and do dramatically modify the properties of the mean-field state, and therefore should in general not be neglected.  For example, if the gauge field is in a confining phase, then the spinons will not be good quasiparticle excitations, as a naive mean-field analysis would suggest -- this indeed occurs in the VCS ground states.  On the other hand, in the CSL and dCSL phases, Chern-Simons terms for the gauge field are present; this not only prevents spinon confinement, it converts the spinons from fermions into anyons.

\section{Properties of topological liquid ground states}
\label{sec:top}

Anticipating the results on energetics discussed below, in this section we discuss the properties of the three topological liquid ground states on the square lattice.  Since the main focus of this paper is on energetics, 
we shall content ourselves primarily with deriving low-energy effective field theories for each state, and shall not discuss the resulting properties in detail.

\subsection{Abelian chiral spin liquid}
\label{sec:acsl}

The Abelian chiral spin liquid (ACSL) occurs for $n_c = 1$, and corresponds to a mean-field saddle point
\begin{eqnarray}
\label{eq:uniformfield}
\bar{\chi}_{\br \br'} &=& \chi e^{i a^0_{\br \br'} } \\
\mu_{\br} &=& 0 \text{,} 
\end{eqnarray}
where $\chi$ is real and positive, and $a^0_{\br \br'}$ is chosen so that $2 \pi / k$ magnetic flux pierces each plaquette of the square lattice.  The band structure consists of $k$ bands, of which the lowest is full and the others are empty, resulting in a Hall conductance (for the mean-field fermions) of  $\sigma_{x y} = N$.

To understand the properties of this state it is necessary to go beyond mean-field level, and couple the fermions to the fluctuating ${\rm U}(1)$ gauge field.  However some properties can already be understood at mean-field level.  In particular, we see that parity (\emph{i.e.} reflection) and time reversal symmetries are spontaneously broken, while the other symmetries of the square lattice (as well as ${\rm SU}(N)$ spin rotation) are preserved.  To see this, it is important to recall that in a slave-particle gauge theory such as this one, symmetry operations act projectively on the fermions.\cite{wen02}  For example, if $S : \br \to S(\br)$ is a space-group operation, then acting on a fermion it may be supplemented by a general space-dependent ${\rm U}(1)$ gauge transformation:
\begin{equation}
S : f_{\br \alpha} \to e^{i \lambda^S_{\br} } f_{ S(\br) \alpha } \text{.}
\end{equation}
An operation $S$ is a symmetry if and only if it is possible to find a gauge transformation $\lambda^S_{\br}$ such that the above transformation leaves the mean-field Hamiltonian invariant.  For the CSL saddle point, this is nothing but the familiar magnetic translation group (expanded to include all symmetries, not only translations).  Because reflections and time reversal both change the sign of the gauge-invariant magnetic flux through each plaquette, they are spontaneously broken in the ACSL.  Other operations leave the flux invariant and are indeed symmetries of the ACSL.

To go beyond mean-field theory, we couple the fermions to the fluctuating ${\rm U}(1)$ gauge field.  Since the fermions are gapped we can integrate them out, resulting in the following imaginary-time continuum effective action for the gauge field:
\begin{equation} \label{eq:chernsimons}
S =  \int d\tau d^2\br \Big[  \frac{i N}{4 \pi}  \epsilon_{\mu \nu \lambda} a_{\mu} \partial_{\nu} a_{\lambda}
+ \frac{1}{2 e^2} ( \sum_{\mu} \epsilon_{\mu \nu \lambda} \partial_{\nu} a_{\lambda} )^2 \Big] \text{.}
\end{equation}
This is simply Maxwell-Chern-Simons theory.  The coefficient of the Chern-Simons term is determined by $\sigma_{x y} = N$, while the coefficient of the Maxwell term is non-universal (in the large-$N$ limit, it is determined by details of the fermion band structure).  Various properties of the ACSL can be derived from this effective action -- notably, it implies that the fermions are converted via flux attachment into anyons with statistics angle $\pi \pm \pi / N$.

A different -- and particularly concrete -- route beyond mean-field theory is to construct a wavefunction for the ACSL.  One starts with the ground state  of the mean-field Hamiltonian, which for the ACSL is simply an integer quantum Hall state with the lowest (lattice) Landau level filled.  One applies the Gutzwiller projection operator ${\cal P}$, which simply projects onto the subspace with exactly $m$ fermions on every lattice site.  By construction, $| \psi \rangle = {\cal P} | \psi_0 \rangle$ satisfies the local constraint Eq.~(\ref{eqn:num-constraint}) and is thus a legitimate wavefunction for the spin model.  Such Gutzwiller projected wavefunctions have been studied and discussed in a variety of contexts (for a few examples, see Refs.~\onlinecite{gros89,wen99,paramekanti01,wen02}), and properties of such wavefunctions can be computed numerically using a Monte Carlo technique.\cite{gros89}  It is reasonable to expect that $| \psi \rangle$ should correctly capture the properties of the corresponding low-energy effective gauge theory; for example, it has been shown that a class of projected wavefunctions associated with an effective $Z_2$ gauge theory  capture the expected $Z_2$ topological order.\cite{ivanov02,paramekanti05}  However, this expectation will need to be tested by future detailed studies of the wavefunction.  In the present case, the projected wavefunction may be a useful tool for future microscopic analysis away from the large-$N$ limit, in particular to help assess the prospects for ACSL in physically realizable models.

Another important property of the ACSL is the presence of gapless chiral edge states, which are described by a chiral ${\rm SU}(N)_1$ Wess-Zumino-Witten (WZW) model.  A simple argument for this can be given following Ref.~\onlinecite{wen91}:  Rather than consider the low-energy effective field theory of fermions coupled to a gauge field, we consider the projected wavefunction described above.  Before projection, the edge mode consists simply of $N$ chiral fermions.  Using non-Abelian bosonization, the edge theory can be cast as two decoupled theories: a chiral ${\rm SU}(N)_1$ Wess-Zumino-Witten (WZW) model, and a chiral ${\rm U}(1)$ Luttinger liquid.\cite{witten84, knizhnik84}  This is an instance of spin-charge separation, where the ${\rm SU}(N)$ spin degrees of freedom are associated with the ${\rm SU}(N)_1$ WZW model, and the fermion ``charge'' with the Luttinger liquid.
Upon projection, the ``charge'' degrees of freedom are removed, and hence so is the ${\rm U}(1)$ Luttinger liquid, while the spin degrees of freedom and the ${\rm SU}(N)_1$ WZW model survive.

Finally, we mention an alternate route to construct a low-energy effective theory for the ACSL that does not require integrating out the fermions.  This approach is based on the Chern-Simons effective theory for Abelian quantum Hall states.\cite{wen95}  In this approach, one pays the price that the ${\rm SU}(N)$ symmetry is broken down to ${\rm U}(1)^{N-1}$, but this is not expected to affect any topological properties of the state.  Before coupling to the gauge field, each spin species of fermion is in an integer quantum Hall state, and the current of the fermions of spin $\alpha$ (where $\alpha = 1,\dots,N$) can be represented in terms of a ${\rm U}(1)$ gauge field:
\begin{equation}
\label{eqn:current-rep-with-gauge-field}
J^{\alpha}_{\mu} = \frac{1}{2\pi} \epsilon_{\mu \nu \lambda} \partial_{\nu} b^\alpha_{\lambda} \text{.}
\end{equation}
The corresponding integer quantum Hall state is captured by a Chern-Simons term for $b^\alpha$, which gives the following contribution to the real-time Lagrangian:
\begin{equation}
{\cal L}_{\alpha} = \frac{1}{4\pi} \epsilon_{\mu \nu \lambda} b^{\alpha}_{\mu} \partial_{\nu} b^{\alpha}_{\lambda} \text{.}
\end{equation}
Moreover, the coupling of the fermions of spin $\alpha$ to the gauge field $a_{\mu}$ is simply given by
\begin{equation}
a_{\mu} J^{\alpha}_{\mu} = \frac{1}{2\pi} \epsilon_{\mu \nu \lambda} a_{\mu} \partial_{\nu} b^{\alpha}_{\lambda} \text{.}
\end{equation}
Finally, the $a_{\mu}$ gauge field has no bare Chern-Simons term -- it is a Lagrange-multiplier field whose role is to make the total ${\rm U}(1)$ fermion current vanish.  (The Chern-Simons term derived above for $a_{\mu}$ came from integrating out the fermions.)  Combining the above results, we have the Lagrangian in $K$-matrix form,
\begin{equation}
{\cal L} = \frac{1}{4\pi} K_{I J} A^I_{\mu} \epsilon_{\mu \nu \lambda} \partial_{\nu} A^J_{\lambda} \text{,}
\end{equation}
where $I = 1,\dots, N+1$, $A^1_{\mu} = a_{\mu}$, $A^I = b^{\alpha - 1}_{\mu}$ for $I > 1$, and the $(N+1)\times(N+1)$ $K$-matrix is
\begin{equation}
K = \left( \begin{array}{cc} 0 & {\cal I}^T \\
{\cal I} & {\bf 1}_{N \times N}
\end{array} \right) \text{.}
\end{equation}
Here, ${\bf 1}_{N \times N}$ is the $N \times N$ identity matrix, and ${\cal I}^T = (1, \dots, 1)$ is a $N$-element vector.  Following Ref.~\onlinecite{wen95}, both bulk and edge topological properties can be deduced from this effective theory.  We note that the $K$-matrix has $N$ positive eigenvalues and one negative eigenvalue, and thus gives rise to $N$ co-propagating edge modes and one counter-propagating mode.  The counter-propagating mode, and one of the co-propagating modes, are singlets under ${\rm U}(1)^{N-1}$ spin rotations, and these singlet modes generically are expected to acquire a gap, leaving $N-1$ gapless co-propagating modes -- this is nothing but the free boson description of the ${\rm SU}(N)_1$ chiral WZW model.

\subsection{Non-Abelian chiral spin liquid}

The non-Abelian chiral spin liquid (nACSL) occurs for $n_c = 2$, and corresponds to a mean-field saddle point
\begin{eqnarray}
\bar{\chi}^{a b}_{\br \br'} &=& \chi e^{i a^0_{\br \br'} } \delta^{a b} \\
\mu^{a b}_{\br} &=& 0 \text{,} 
\end{eqnarray}
where $\chi$ is real and positive, and again $a^0_{\br \br'}$ is chosen so that $2 \pi / k$ magnetic flux pierces each plaquette of the square lattice.  This state has a ${\rm U}(2) = {\rm U}(1) \times {\rm SU}(2)$ gauge structure, and upon going beyond mean-field theory the fermions are coupled to a ${\rm U}(2)$ gauge field.  The background magnetic flux is a ${\rm U}(1)$ flux -- the background ${\rm SU}(2)$ flux is zero.
The band structure can be thought of as $k$ $2N$-fold degenerate bands, where the lowest band is filled and all others are empty.  The mean-field fermions have a Hall conductance $\sigma_{x y} = 2 N$.  As above, parity and time reversal are spontaneously broken, while other symmetries are preserved.

In the large-$N$ limit, the ground state energy of the nACSL is precisely twice that of the ACSL.  This occurs because, at the mean-field level, the nACSL is simply two decoupled copies of the $n_c = 1$ ACSL, each with the same magnetic flux.

Upon integrating out the fermions, we obtain the following action:
\begin{eqnarray}
S &=& \frac{2 N i}{4 \pi} \int d\tau d^2 \br \, \epsilon_{\mu \nu \lambda} a_{\mu} \partial_{\nu} a_{\lambda} \nonumber \\ &+& \frac{i N}{4 \pi} \int d\tau d^2 \br \,  \epsilon_{\mu \nu \lambda} \operatorname{tr} \Big[  \alpha_{\mu} \partial_{\nu} \alpha_{\lambda} - \frac{ 2 i }{3} \alpha_{\mu} \alpha_{\nu} \alpha_{\lambda} \Big] \text{.} 
\end{eqnarray}
Here $a_{\mu}$ is the ${\rm U}(1)$ gauge field, $\alpha_{\mu} = \sum_{i = 1}^{3} \alpha^i_{\mu} \sigma^i$ is the ${\rm SU}(2)$ gauge field ($\sigma^i$ are the usual Pauli matrices), and we omitted the Maxwell terms that are also present.  The second term is the level-$N$ Chern-Simons term for the ${\rm SU}(2)$ gauge field, which gives rise to the non-Abelian statistics of the nACSL.

As above for the ACSL, one can construct a wavefunction for the nACSL.  One proceeds as above, but now must apply a projection operator to enforce both the local constraints Eq.~(\ref{eqn:num-constraint}) and Eq.~(\ref{eqn:color-constraint}).  

The chiral edge states of the nACSL can be understood in terms of an argument very similar to that given above for the ACSL.  In mean-field theory, there are $2N$ chiral fermions on the edge of the system.  Following Affleck,\cite{affleck86} this free fermion theory can be bosonized to a chiral ${\rm SU}(N)_2$ WZW model (carrying spin excitations), a chiral ${\rm SU}(2)_N$ WZW model (carrying color), and a chiral ${\rm U}(1)$ Luttinger liquid.  Now the projection removes both the ``charge'' and color degrees of freedom of the fermions, leaving only the chiral ${\rm SU}(N)_2$ WZW model.

\subsection{Doubled chiral spin liquid}
\label{sec:dcsl}

The doubled chiral spin liquid (dCSL) occurs for $n_c = 2$, and corresponds to a mean-field saddle point
\begin{eqnarray}
\bar{\chi}^{a b}_{\br \br'} &=& \left( \begin{array}{cc}
e^{i a^0_{\br \br'}} & 0 \\
0 & e^{-i a^0_{\br \br'}} \end{array}\right) \label{eqn:dcsl-chi} \\
\mu^{a b}_{\br} &=& 0 \text{,} 
\end{eqnarray}
where $\chi$ and $a^0_{\br \br'}$ are as above.  In contrast to the nACSL, there is now a ${\rm SU}(2)$ background magnetic flux, but no ${\rm U}(1)$ flux.  Following the reasoning of Ref.~\onlinecite{wen91b}, the presence of the nontrivial ${\rm SU}(2)$ flux breaks the ${\rm SU}(2)$ gauge structure down to ${\rm U}(1)$. More precisely, the $\alpha^1$ and $\alpha^2$ components of the ${\rm SU}(2)$ gauge field acquire a mass due to the presence of the flux, while the $\alpha^3$ component is unaffected.  Therefore, for the purposes of understanding the low-energy physics, we can drop the $\alpha^1$ and $\alpha^2$ components of the ${\rm SU}(2)$ gauge field, and consider a theory of fermions coupled to the two ${\rm U}(1)$ gauge fields $a_{\mu}$ and $\alpha^3_{\mu}$.  It should be noted that the special role of $\alpha^3_{\mu}$, as compared to $\alpha^1_{\mu}$ and  $\alpha^2_{\mu}$, is determined by the choice of gauge made in writing Eq.~(\ref{eqn:dcsl-chi}) -- a global ${\rm SU}(2)$ gauge transformation can be made to select any desired preferred axis.

In the large-$N$ limit, the ground state energy of the dCSL is again precisely twice that of the ACSL, because again, at the mean-field level, the dCSL is two decoupled copies of the $n_c = 1$ ACSL, but now with opposite magnetic fluxes.  This means that in the $N \to \infty$ limit the dCSL and nACSL have exactly the same energy.  This degeneracy is expected to be lifted by $1/N$ corrections that can in principle be computed; this is left for future work.

The dCSL actually respects time reversal symmetry, which is implemented by the operation
\begin{equation}
{\cal T} : f_{\br a \alpha} \to (i \sigma^2)_{a b} f_{\br b \alpha} \text{.}
\end{equation}
(This operation can be supplemented as well with a ${\rm SU}(N)$ rotation, but due to the ${\rm SU}(N)$ symmetry this is not essential.)  The crucial point is that the gauge-rotation in the color space compensates for the fact that complex conjugation reverses the flux.  Reflection symmetry ${\cal R} : \br \to \br'$, where $\br' = (-r_x, r_y)$, is similarly preserved, and
\begin{equation}
{\cal R} : f_{\br a \alpha} \to (i \sigma^2)_{a b} f_{\br' b \alpha} \text{.}
\end{equation}
The other symmetries (lattice translations and rotations, and ${\rm SU}(N)$ spin rotations) are preserved in the dCSL as they are in the above two states.  The dCSL therefore does not spontaneously break \emph{any} symmetries, in contrast to the ACSL and nACSL.

Upon integrating out the fermions, we obtain the following mutual Chern-Simons action:
\begin{equation}
S = \frac{i N}{\pi} \int d\tau d^2\br \, \epsilon_{\mu \nu \lambda} \, a_{\mu} \partial_{\nu} \alpha^3_{\lambda} \text{.}
\end{equation}
Here we have again omitted the Maxwell terms that will also be present; the mutual Chern-Simons term fully gaps out both gauge fields, and the Maxwell terms play only the quantitative role of setting the scale of the gap to gauge field excitations.  It should be noted that similar spin liquid states, but with an additional non-Abelian gauge structure, were considered in Ref.~\onlinecite{freedman04}.  (There, however, the analog of the $\alpha^3_{\mu}$ gauge field was incorrectly dropped, and therefore a ${\rm U}(1)$ mutual Chern-Simons term was missed.)  It can be seen that this term also converts the mean-field fermionic excitations into anyons with statistics angle $\pi \pm \pi / N$, which can occur in a time-reversal invariant fashion due to the color index.  We note that the same procedure described for the nACSL can be applied here to produce a wavefunction for the dCSL.

Because the dCSL respects time reversal symmetry, it lacks chiral edge states.  However, it is interesting to note that -- when $N$ is odd -- the edge states are protected at the mean-field level, because the mean-field Hamiltonian has a nontrivial $Z_2$ topological invariant\cite{kane05b} for odd $N$.  It is therefore conceivable that topologically protected edge states could survive coupling of the mean-field fermions to the fluctuating gauge fields, and it would be interesting to study this question.  Presumably such protection, if it occurs, would only hold if one assumes that no spontaneous breaking of time-reversal symmetry occurs at the edge.

Finally, we can also construct a $K$-matrix Lagrangian for the dCSL as above for the ACSL.  We let $A^1_{\mu} = a_{\mu}$ and $A^2_{\mu} = \alpha^3_{\mu}$.  Next, for $I = 3, \dots, N + 2$, $A^I_{mu}$ represents the current of fermions with $a = 1$ and spin $\alpha = I - 2$ (as in Eq.~\ref{eqn:current-rep-with-gauge-field}), while for $I = N+3, \dots, 2 N + 2$, $A^I_{\mu}$ represents the current of fermions with $a = 2$ and spin $\alpha = I - (N+2)$.  Following essentially the same reasoning as in Sec.~\ref{sec:acsl}, we have the $(2 N + 2) \times (2 N + 2)$ $K$-matrix
\begin{equation}
K = \left( \begin{array}{cccc}
0 & 0 & {\cal I}^T & {\cal I}^T \\
0 & 0 & {\cal I}^T & - {\cal I}^T \\
{\cal I} & {\cal I} & {\bf 1}_{N \times N} & {\bf 0}_{N \times N} \\
{\cal I} & - {\cal I} & {\bf 0}_{N \times N} & - {\bf 1}_{N \times N}
\end{array} \right) \text{,}
\end{equation}
where ${\bf 0}_{N \times N}$ is the $N \times N$ matrix of zeros.

\section{Large-$N$ limit: General lattices}
\label{sec:genlat}

In Ref.~\onlinecite{rokhsar90}, Rokhsar derived an exact lower bound on the large-$N$ ground state energy $E_{{\rm MFT}}$, for the case $n_c = 1$ and $k = 2$.  He further showed that this bound is saturated by a VBS state under conditions that are satisfied for the great majority of lattices one encounters.  More precisely, let us define ${\cal J}_{{\rm max}}$ to be the largest of the exchange couplings ${\cal J}_{\br \br'}$.  (Note that we do \emph{not} restrict to only nearest-neighbor exchange.)  Following Rokhsar we say that a lattice is dimerizable with respect to ${\cal J}_{{\rm max}}$ when it is possible to partition the lattice into 2-site dimers, such that the two sites $(\br, \br')$ in each dimer have ${\cal J}_{\br \br'} = {\cal J}_{{\rm max}}$.  Each lattice site must belong to precisely one dimer.  For a fixed partition into dimers, let the set of bonds $(\br, \br')$ that connect the two sites of a dimer be $B$.  Rokhsar considered the VBS saddle point defined by
\begin{eqnarray}
\begin{array}{ll} \chi_{\br \br'} = \chi \neq 0 & \text{, } (\br, \br') \in B \\
\chi_{\br \br'} = 0 & \text{, } (\br, \br') \notin B 
\end{array} \text{,}
\end{eqnarray}
and showed that it is a ground state (its energy saturates the bound on $E_{{\rm MFT}}$).  Except in the case of disordered systems lacking translation symmetry, most familiar lattices (and associated sets of ${\cal J}_{\br \br'}$) are dimerizable with respect to ${\cal J}_{{\rm max}}$.\cite{rokhsar90}  Therefore, when $k=2$, one has to consider a relatively unusual lattice to find anything other than a VBS ground state in the large-$N$ limit.  (See Ref.~\onlinecite{rokhsar90} for an example of a lattice that is not dimerizable with respect to ${\cal J}_{{\rm max}}$.)

Here, we generalize Rokhsar's bound to the case of arbitrary $k$ and $n_c$ (Sec.~\ref{sec:bound}).  In Sec.~\ref{sec:saturation} we derive necessary and sufficient conditions to saturate the bound.  Next, in Sec.~\ref{sec:simplex}, we show that the analog of Rokhsar's VBS saddle point is a $k$-simplex VCS state, where the lattice is decomposed into $k$-site simplices ($k$-simplices for short), in which every site is connected to the other $k-1$ sites by an exchange coupling ${\cal J}_{\br \br'} = {\cal J}_{{\rm max}}$.  As soon as $k > 2$, \emph{many} lattices cannot be decomposed into $k$-simplices, and for $k \geq 5$ we show that no lattice can be decomposed into $k$-simplices without fine-tuning of the exchange couplings.  Therefore it becomes more and more difficult to saturate the bound as $k$ increases.

\subsection{Derivation of the bound}
\label{sec:bound}

Our starting point is the mean-field Hamiltonian ${\cal H}_{{\rm MFT}}$ for general $n_c$  [Eq.~(\ref{eqn:hmft})],   where $\bar{\chi}^{a b}_{\br \br'}$ and $\mu^{a b}_{\br}$ are chosen to satisfy the saddle-point equations Eqs.~(\ref{eqn:spe-chi},\ref{eqn:spe-constraints}). 
The bound is derived in two steps:  first we will show $E_{{\rm MFT}} \geq E'_{{\rm MFT}}$ (defined below), then we will show $E'_{{\rm MFT}} \geq E_{{\rm bound}}$.  The first step was actually omitted in Ref.~\onlinecite{rokhsar90}.  While this step should not be omitted even in the special case considered there, none of the results of Ref.~\onlinecite{rokhsar90} are affected by this omission.

Recalling the definitions of ${\cal H}_{{\rm MFT}}$ in Eq.(\ref{eqn:hmft}) and ${\cal H}_K$ in Eq.~(\ref{eqn:hk}), we begin by defining
\begin{equation}
{\cal H}'_{{\rm MFT}} = \prsum \frac{N}{{\cal J}_{\br \br'} } \operatorname{tr} ( \bar{\chi}^\dagger_{\br \br'} \bar{\chi}^{\vphantom\dagger}_{\br \br'} )  + {\cal H}_K \text{,}
\end{equation}
where $\bar{\chi}^{a b}_{\br \br'}$ is the same as in ${\cal H}_{{\rm MFT}}$.   
 That is, we obtain ${\cal H}'_{{\rm MFT}}$ by starting with ${\cal H}_{{\rm MFT}}$ and setting $\mu^{a b}_{\br}$ to zero.  The ground state energy of ${\cal H}'_{{\rm MFT}}$ is $E'_{{\rm MFT}}$.  Note that, in general, the ground state of ${\cal H}'_{{\rm MFT}}$ will not satisfy the saddle point equations.

Now, $E_{{\rm MFT}} = \langle {\cal H}_{{\rm MFT}} \rangle$, where the expectation value is taken using the ground state of ${\cal H}_{{\rm MFT}}$.  Using Eq.~(\ref{eqn:spe-constraints}), we note that $\langle {\cal H}_V \rangle = - m \sum_{\br}\operatorname{tr} (\mu_{\br}) $; this cancels the second term in ${\cal H}_{{\rm MFT}}$, so we have
\begin{equation}
E_{{\rm MFT}} =  \prsum \frac{N}{{\cal J}_{\br \br'} } \operatorname{tr} ( \bar{\chi}^\dagger_{\br \br'} \bar{\chi}^{\vphantom\dagger}_{\br \br'} ) + \langle {\cal H}_{K} \rangle \text{.}
\end{equation}
Letting $E_{K}$ be the ground state energy of ${\cal H}_K$, we have $\langle {\cal H}_K \rangle \geq E_{K}$, and so
\begin{equation}
E_{{\rm MFT}} \geq E'_{{\rm MFT}} = N \prsum \frac{| \chi_{\br \br'}|^2}{J_{\br \br'} } + E_{K} \text{.}
\end{equation}
This is the first of the two desired inequalities.

We shall now deal with ${\cal H}'_{{\rm MFT}}$ and $E'_{{\rm MFT}}$, and establish a lower bound on $E'_{{\rm MFT}}$.  To do this, we generalize Rokhsar's argument\cite{rokhsar90} to the case of general $m$ and $n_c$.   Let $N_s$ be the number of sites of our lattice.  We label the single-particle energy levels of ${\cal H}_K$ by an index $q$; the energies are $\epsilon_q$.  ${\cal H}_K$ is specified by the $n_c N N_s \times n_c N N_s$ Hermitian matrix
\begin{equation}
(H_K)_{\br a \alpha ; \br' b \beta} = \delta_{\alpha \beta} \chi^{a b}_{\br \br'} \text{,}
\end{equation}
where
\begin{equation}
\chi_{\br' \br} = \chi^\dagger_{\br \br'} \text{.}
\end{equation}
Because this is traceless (all diagonal entries are zero), we have
\begin{equation}
\sum_q \epsilon_q = 0 \text{.} \label{eqn:sum-to-zero}
\end{equation}  The ground state of ${\cal H}_{K}$ (and hence of ${\cal H}'_{{\rm MFT}}$) is obtained by filling the lowest $n_c m N_s$ energy levels with fermions.  We call the set of such energy levels ${\cal L}$.  The other $n_c(N-m)N_s$ levels, which we denote by the set ${\cal U}$, are empty.

It will be useful to define averages over the sets of levels ${\cal L}$ and ${\cal U}$:
\begin{eqnarray}
\left[ \epsilon \right]_{\cal L} &=& \frac{1}{n_c m N_s} \sum_{q \in {\cal L}} \epsilon_{q} \\
\left[ \epsilon \right]_{\cal U} &=& \frac{1}{n_c (N - m) N_s} \sum_{q \in {\cal U}} \epsilon_q \text{.}
\end{eqnarray}
We also denote the average of $\epsilon^2_q$ over the two sets by $[ \epsilon^2]_{\cal L}$ and $[ \epsilon^2 ]_{\cal U}$, and the average of $\epsilon^2_q$ over \emph{all} states is written $[ \epsilon^2 ]$.
Equation~(\ref{eqn:sum-to-zero}) implies
\begin{equation}
\label{eqn:zerotrace}
[ \epsilon ]_{\cal U} = - \frac{m}{N - m} [ \epsilon ]_{\cal L} \text{.}
\end{equation}

The bound originates from the pair of inequalities
\begin{eqnarray}
\label{eqn:seL}
[ \epsilon ]_{\cal L}^2 &\leq& [ \epsilon^2 ]_{\cal L}  \\
\label{eqn:seU}
[ \epsilon ]_{\cal U}^2 &\leq& [ \epsilon^2 ]_{\cal U} \text{,}
\end{eqnarray}
which just express the fact that variance is positive.  These inequalities are saturated (become equalities) if and only if $\epsilon_q$ is constant over each of the sets ${\cal L}$ and ${\cal U}$.
Multiplying Eq.~(\ref{eqn:seL}) by $m/N$, Eq.~(\ref{eqn:seU}) by $(N-m)/N$, and adding the two, we have
\begin{equation}
\frac{m}{N} [ \epsilon ]_{\cal L}^2 + \frac{(N-m)}{N} [ \epsilon ]_{\cal U}^2 \leq [ \epsilon^2 ] \text{.}
\end{equation}
Using Eq.~(\ref{eqn:zerotrace}) and the fact that $[ \epsilon]_{\cal L} < 0$, we have
\begin{equation}
\label{eqn:Laverage}
[\epsilon]_{\cal L} \geq - \sqrt{\frac{N-m}{m}} \sqrt{ [\epsilon^2 ] } \text{.}
\end{equation}
Now, $E_{K} = n_c m N_s [\epsilon]_{\cal L}$, so we have shown
\begin{equation}
E_{K} \geq - n_c m N_s \sqrt{\frac{N-m}{m}} \sqrt{[\epsilon^2]} \text{.}
\end{equation}

The next step is to get a simple expression for $[\epsilon^2]$.  We have
\begin{eqnarray}
[\epsilon^2] &=& \frac{1}{n_c N N_s} \sum_{\alpha} \epsilon^2_{\alpha}
= \frac{1}{n_c N N_s} \operatorname{tr} [ H_K^2 ] \nonumber \\
&=&  \frac{2}{n_c N_s} \sum_{(\br, \br')} \operatorname{tr} ( \chi^\dagger_{\br \br'} \chi^{\vphantom\dagger}_{\br \br'} ) \text{.}
\end{eqnarray}
Therefore we have the inequality
\begin{eqnarray}
E'_{{\rm MFT}} &\geq& N \prsum \sum_{a, b} \frac{| \chi^{a b}_{\br \br'}|^2}{{\cal J}_{\br \br'} } \label{eqn:non-min-bound} \\
&-& n_c m N_s \sqrt{\frac{N-m}{m}} \sqrt{ \frac{2}{n_c N_s} \sum_{(\br, \br')} \sum_{a, b} | \chi^{a b}_{\br \br'}|^2 }  \text{.} \nonumber
\end{eqnarray}

The next step is to minimize this lower bound, which we do by taking the derivative of the right-hand side of Eq.~(\ref{eqn:non-min-bound}) with respect to $| \chi^{a b}_{\br \br'}|$ and setting it to zero:
\begin{equation}
0 = 2 N \frac{|\chi^{a b}_{\br \br'}|}{{\cal J}_{\br \br'}} - \frac{ 2 m \sqrt{(N-m)/m} | \chi^{a b}_{\br \br'}| }{\sqrt{ \frac{2}{n_c N_s} \sum_{(\br'', \br''')} \sum_{c,d} | \chi^{c d}_{\br'' \br'''}|^2 } } \text{.}
\end{equation}
For a given bond $(\br, \br')$, this equation implies that either $|\chi^{a b}_{\br \br'} | = 0$ for all $a,b$, or
\begin{equation}
\label{eqn:mineq}
\frac{2}{n_c N_s} \sum_{( \br'', \br''')} \sum_{c, d} | \chi^{c d}_{\br'' \br'''}|^2 = \frac{m (N-m)}{N^2} {\cal J}_{\br \br'}^2 \text{.}
\end{equation}
Now, the left-hand side of Eq.~(\ref{eqn:mineq}) is  independent of the bond $(\br , \br')$, and so we must have
\begin{equation}
\frac{2}{n_c N_s} \sum_{( \br'', \br''')} \sum_{c, d} | \chi^{c d}_{\br'' \br'''}|^2 =  \frac{m (N-m)}{N^2} {\cal J}^2_{*}
\end{equation}
for some constant ${\cal J}_{*}$.  Moreover, this implies that, for a given bond, unless ${\cal J}_{\br \br'} = {\cal J}_{*}$, then we must have $\chi^{a b}_{\br \br'} = 0$ (for all $a,b$).  Therefore
\begin{eqnarray}
\prsum \sum_{a,b} \frac{| \chi^{a b}_{\br \br'}|^2}{{\cal J}_{\br \br'} } &=& \frac{1}{{\cal J}_{*}} \sum_{(\br, \br')} \sum_{a,b} | \chi^{a b}_{\br \br'} |^2 \\
&=& \frac{n_c N_s}{2} \frac{m (N-m)}{N^2} {\cal J}_{*} \text{.}
\end{eqnarray}

Putting these results into Eq.~(\ref{eqn:non-min-bound}), we have
\begin{equation}
E'_{{\rm MFT}} \geq - \frac{n_c N_s}{2} \frac{m (N-m)}{N} {\cal J}_{*} \text{.}
\end{equation}
The global minimum is clearly achieved when ${\cal J}_{*} = {\cal J}_{{\rm max}}$, the largest of the ${\cal J}_{\br \br'}$.  Therefore
\begin{equation}
\label{eqn:bound}
E_{{\rm MFT}} \geq E'_{{\rm MFT}} \geq - \frac{n_c N_s}{2} \frac{m (N- m)}{N} {\cal J}_{{\rm max}} \text{.}
\end{equation}
Putting $m = N/k$ we have
\begin{equation}
\label{eqn:rokhsar-bound}
E_{{\rm MFT}} \geq  - n_c N N_s \frac{ (k-1)}{2 k^2} {\cal J}_{{\rm max}} \text{,}
\end{equation}
which reduces to Rokhsar's result when $k = 2$ and $n_c = 1$.

\subsection{Necessary and sufficient conditions to saturate the bound}
\label{sec:saturation}

Here, we show that the bound Eq.~(\ref{eqn:rokhsar-bound}) is saturated if and only if the following two conditions hold:  (1)  $\epsilon_q$ is constant over each of the sets ${\cal L}$ and ${\cal U}$.  That is, all the filled states have the same energy, and all empty states have the same energy.  (2) The color density $\tilde{n}^{a b}_{\br}$ calculated using ${\cal H}_K$ satisfies the condition
\begin{equation}
\sum_{\br} \operatorname{tr} (\mu_{\br} \tilde{n}_{\br} ) = 0 \text{.}
\end{equation}
This color density is defined by
\begin{equation}
\tilde{n}^{a b}_{\br} = \langle \hat{n}^{a b}_{\br} \rangle_K \text{,}
\end{equation}
where the expectation value is taken using the ground state of ${\cal H}_K$.  Note that $\tilde{n}_{\br}$ in general does not satisfy the constraint Eq.~(\ref{eqn:spe-constraints}).  These conditions for saturation are very restrictive, as we discuss below.

There are two separate inequalities that must both be turned into equalities for the bound to be saturated.  
The first is $E'_{{\rm MFT}} \geq E_{{\rm bound}}$, and the second is $E_{{\rm MFT}} \geq E'_{{\rm MFT}}$.  Saturation of the first and second inequalities leads to conditions (1) and (2) above, respectively.  It is trivial to show that the first inequality is saturated if and only if $\epsilon_q$ is constant over each of the sets ${\cal L}$ and ${\cal U}$.  

We now show that condition (2) is equivalent to saturation of the second inequality.  It will be useful to define a continuous family of Hamiltonians parametrized by $\alpha \in [0, 1]$:
\begin{equation}
{\cal H}_{\alpha} = {\cal H}_K + \alpha {\cal H}_V \text{.}
\end{equation}
This interpolates between ${\cal H}_K$ at $\alpha = 0$ and ${\cal H}_K + {\cal H}_V$, the fermionic part of ${\cal H}_{{\rm MFT}}$, at $\alpha = 1$.
The ground state of ${\cal H}_{\alpha}$ with energy $E_{\alpha}$ is denoted by $| \psi_{\alpha} \rangle$.  Because we work in the canonical ensemble for the fermion number, we are free to make a constant shift $\mu^{a b}_{\br} \to \mu^{a b}_{\br} + c \delta^{a b}$ so that $\sum_{\br} \operatorname{tr}  (\mu_{\br} )  = 0$ (note that this shift does not change $E_{{\rm MFT}}$).  
With this choice for $\mu_{\br}$, we have
\begin{equation}
\langle \psi_1 | {\cal H}_V | \psi_1 \rangle = 0 \text{.}
\end{equation}
We also have $E_{{\rm MFT}}  - E'_{{\rm MFT}} = E_1 - E_0$.  In particular, $E_{{\rm MFT}} = E'_{{\rm MFT}}$ if and only if $E_0 = E_1$.

The variational principle implies  $\langle \psi_{\alpha} | {\cal H}_{\alpha'} | \psi_{\alpha} \rangle \geq E_{\alpha'}$.  The left-hand side of this inequality can be written
\begin{equation}
\langle \psi_{\alpha} | {\cal H}_{\alpha'} | \psi_{\alpha} \rangle = E_{\alpha} +  (\alpha' - \alpha) \langle \psi_{\alpha} | {\cal H}_V | \psi_{\alpha} \rangle \text{.}
\end{equation}
We have thus shown
\begin{equation}
E_{\alpha} +  (\alpha' - \alpha) \langle \psi_{\alpha} | {\cal H}_V | \psi_{\alpha} \rangle \geq E_{\alpha'} \text{.}
\end{equation}
If we put $\alpha = 1$, this gives $E_1 \geq E_{\alpha}$. 
On the other hand, putting $\alpha' = 1$ gives instead
\begin{equation}
E_{\alpha} + (1 - \alpha) \langle \psi_{\alpha} | {\cal H}_V | \psi_{\alpha} \rangle \geq E_1 \text{.}
\end{equation}
Combining these together,
\begin{equation}
\label{eqn:ineq1}
E_{\alpha} + (1 - \alpha) \langle \psi_{\alpha} | {\cal H}_V | \psi_{\alpha} \rangle \geq E_1 \geq E_{\alpha} \text{,}
\end{equation}
which immediately implies
\begin{equation}
 \langle \psi_{\alpha} | {\cal H}_V | \psi_{\alpha} \rangle \geq 0 \text{.}
 \end{equation}

 A special case of Eq.~(\ref{eqn:ineq1}) is
 \begin{equation}
 E_0 + \langle \psi_0 | {\cal H}_V | \psi_0 \rangle \geq E_1 \geq E_0 \text{.}
 \end{equation}
 From this it follows that if $\langle \psi_0 | {\cal H}_V | \psi_0 \rangle = 0$, then $E_1 = E_0$.  Now suppose the converse, \emph{i.e.} suppose $E_0 = E_1$.  Note that first-order perturbation theory gives us
\begin{equation}
\frac{d E}{d \alpha} = \langle \psi_{\alpha} | {\cal H}_V | \psi_{\alpha} \rangle \text{,}
\end{equation}
and so
\begin{equation}
E_1 - E_0 = \int_0^1 d\alpha  \langle \psi_{\alpha} | {\cal H}_V | \psi_{\alpha} \rangle \text{.}
\end{equation}
By assumption this integral is equal to zero.  Since the integrand is nonnegative, then we must have
$\langle \psi_{\alpha} | {\cal H}_V | \psi_{\alpha} \rangle = 0$, and in particular for $\alpha = 0$.  

Therefore we have shown that $E_1 = E_0$ if and only if $\langle \psi_0 | {\cal H}_V | \psi_0 \rangle = 0$, and hence $E_{{\rm MFT}} = E'_{{\rm MFT}}$ if and only if $\langle \psi_0 | {\cal H}_V | \psi_0 \rangle = 0$.  Since $\langle \psi_0 | {\cal H}_V | \psi_0 \rangle = \sum_{\br} \operatorname{tr} ( \mu_{\br} \tilde{n}_{\br} )$, we have established condition (2) as desired.

Both conditions derived above for  saturation of the bound are highly restrictive.  Condition (1) dictates that there be only two energies in the spectrum; we should expect this to occur only when $\chi^{a b}_{\br \br'}$ is such that the lattice is broken into clusters, so that the spectrum consists of perfectly flat bands.  Condition (2) is also very restrictive.  An easy way to satisfy (2) is simply to have a saddle point where $\mu_{\br} = 0$.  Suppose instead that $\mu_{\br} \neq 0$, and so generically we should expect that $\tilde{n}_{\br}$ is non-uniform and does not satisfy Eq.~(\ref{eqn:spe-constraints}).  It is useful to imagine starting from $\alpha = 0$ and turning on ${\cal H}_V$ by increasing $\alpha$.  The $\mu_{\br}$ need to be chosen to ``even out'' the color density, so that it satisfies Eq.~(\ref{eqn:spe-constraints}) once $\alpha = 1$.  Naively, a choice of $\mu_{\br}$ accomplishing this will cost energy at each lattice site; that is,
\begin{equation}
\langle \psi_0 | \big[ - \operatorname{tr} (\mu_{\br} \hat{n}_{\br} ) \big] | \psi_0 \rangle > 0 \text{.}
\end{equation}
This would imply
\begin{equation}
\sum_{\br} \operatorname{tr} ( \mu_{\br} \tilde{n}_{\br} ) < 0 \text{,}
\end{equation}
which is in conflict with condition (2).  This discussion indicates that satisfying condition (2) when $\mu_{\br} \neq 0$ is unlikely.

\subsection{Saturation of the bound and $k$-simplex VCS states}
\label{sec:simplex}

The necessary and sufficient conditions derived above still leave open the questions of what kind of saddle points saturate the bound,
and whether saturation is possible for a given lattice and set of exchange couplings $J_{\br \br'}$.  Saturation is not always possible -- for example, on any bipartite lattice with $k > 2$, the stricter bound derived in Sec.~\ref{sec:bipartite} shows that saturation of Eq.~(\ref{eqn:rokhsar-bound}) is impossible.  Here, we will show that, when they exist,  $k$-simplex VCS states saturate the bound and are thus the analogs of the VBS states for $k = 2$.
In striking contrast to VBS states, many commonly encountered lattices do not admit any $k$-simplex VCS states for $k > 2$.  Moreover, for $k > d+1$ there is no $d$-dimensional lattice that admits a $k$-simplex state without fine-tuning of the exchange couplings.  The implication is that for $k > 2$ a much wider range of ground states are possible in the large-$N$ limit, including spin liquid states.  Unless stated otherwise, when discussing specific lattices we consider the case of nearest-neighbor exchange only.

\begin{figure}
\includegraphics[width=3in]{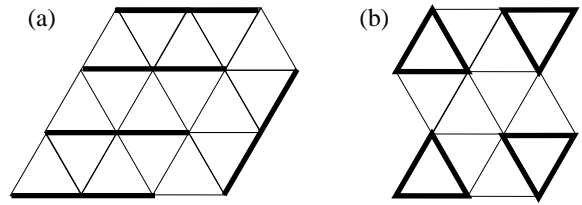}
\caption{Illustration of 3-cluster and 3-simplex VCS states on the triangular lattice (for $n_c = 1$).  $\chi_{\br \br'}$ is nonzero on the highlighted bonds and zero elsewhere.  Both states (a) and (b) are 3-cluster states.  State (b) is a 3-simplex state and is a $N = \infty$ ground state of the $k=3$ triangular lattice model.  State (a) is not a 3-simplex state and therefore has higher energy than (b) following the discussion in the text.}
\label{fig:tri}
\end{figure}

We shall first discuss $k$-simplex VCS states for the simpler case $n_c = 1$, and then generalize to arbitrary $n_c$.  In the large-$N$ limit, by VCS state we mean a saddle point where $\chi_{\br \br'}$ is chosen to decompose the lattice into clusters.  Each lattice site belongs to exactly one cluster, and any two sites in the same cluster are connected by $\chi_{\br \br'} \neq 0$ along some path of bonds (they need not be directly connected).  Each cluster must contain some multiple of $k$ lattice sites, since otherwise the cluster will not be a singlet.  In a $k$-cluster state, every cluster contains exactly $k$ sites.  A $k$-simplex state is a $k$-cluster state where, within each cluster, each site is (directly) connected to every other site by a single bond with ${\cal J}_{\br \br'} = {\cal J}_{{\rm max}}$ (see Fig.~\ref{fig:tri}).  Just as for VBS states, on a given lattice there can be many different $k$-simplex states with the same $N \to \infty$ energy.  It is expected that $1/N$ corrections will select a particular ordered pattern out of this degenerate manifold, again precisely as for VBS states.\cite{read89b}

\begin{figure}
\includegraphics[width=1.6in]{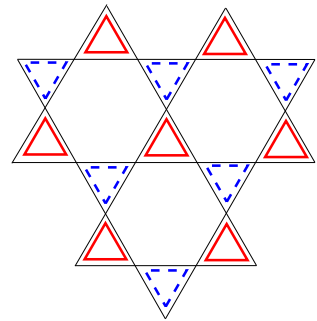}
\caption{An example $k$-simplex state with $n_c = 2$ and $k=3$, on the kagome lattice.  Simplices of one color are the triangles marked with solid lines (red online), and those of the other color are triangles marked with dashed lines (blue online).  This state was discussed (for $N=3$) in Ref.~\onlinecite{arovas08}.}
\label{fig:kag}
\end{figure}

To generalize $k$-cluster states to $n_c > 1$, we consider only \emph{diagonal} $\chi^{a b}_{\br \br'}$ (and $\mu^{a b}_{\br}$).  That is, we consider
\begin{eqnarray}
\label{eqn:diagonal}
\chi^{a b}_{\br \br'} &=& \delta^{a b} \chi^a_{\br \br'} \text{   (no sum).} \\
\mu^{a b}_{\br} &=& \delta^{a b} \mu^a_{\br} \text{    (no sum).}
\end{eqnarray}
For each $a = 1,\dots,n_c$, $\chi^a_{\br \br'}$ is chosen to give a $k$-cluster decomposition of the lattice, resulting in $n_c$ different $k$-cluster decompositions.  A $k$-simplex state occurs where each $k$-cluster decomposition is also a decomposition into $k$-simplices; an example of a  $n_c > 1$ $k$-simplex states is given in Fig.~\ref{fig:kag}  Such states were considered 
for $k = 2$ in Ref.~\onlinecite{read89b}, and also as exact ground states of special models for a variety of $n_c$ and $N$ in Ref.~\onlinecite{arovas08}.

Focusing on a single color (say, $a =1$) and a single cluster, and choosing $\mu^{a b}_{\br} = 0$ and $\chi^1_{\br \br'} \to -\chi$ (for bonds within a cluster), the fermionic part of the mean-field Hamiltonian 
 in a $k$-simplex state is
\begin{equation}
{\cal H}^{k-{\rm simplex}}_{F} =  - \chi \sum_{\br \neq \br'} f^\dagger_{\br 1 \alpha} f^{\vphantom\dagger}_{\br' 1 \alpha} \text{.}
\end{equation}
The lowest single-particle energy is $\epsilon_{{\cal L}} = -(k-1)\chi$; the $k-1$ other eigenvalues are degenerate and take the value $\epsilon_{{\cal U}} = \chi$.  The ground state is obtained by filling the lowest level in all clusters, and it is easy to see that in this state the saddle-point condition
$\langle \hat{n}^{a b}_{\br} \rangle = m \delta^{a b}$ is satisfied.  This state satisfies both the conditions for saturation of the lower bound Eq.~(\ref{eqn:rokhsar-bound}), and this is easily verified by direct computation of the energy.  

More general $k$-cluster states do not saturate the bound. To illustrate this, consider for simplicity $n_c = 1$ and a lattice where either ${\cal J}_{\br \br'} = {\cal J}_{{\rm max}}$, or ${\cal J}_{\br \br'} = 0$.  Consider a $k$-cluster state where all the clusters are identical and each cluster contains $N_b$ bonds with nonzero exchange.  It can be shown that the energy of each cluster is $E_c = - N N_b {\cal J}_{{\rm max}} / k^2$ (Appendix~\ref{app:kcluster}), so the total energy is then
\begin{equation}
E_{{\rm MFT}} = \frac{N_s}{k} E_c = - \frac{-N {\cal J}_{{\rm max}} N_s N_b}{k^3} \text{.}
\end{equation}
This attains the bound only if the number of bonds is maximum, that is $N_b = k (k-1)/2$ -- but this is precisely the condition that each cluster is a $k$-simplex.

As mentioned above, while most lattices admit a VBS state, this is not the case for $k$-simplex states with $k > 2$.  For example, the
square and honeycomb lattices admit VBS states but no $k$-simplex states with $k \geq 3$.  The triangular (Fig.~\ref{fig:tri}) and kagome lattices admit both VBS and 3-simplex states, but lack $k$-simplex states for $k \geq 4$.  The three-dimensional pyrochlore lattice of corner-sharing tetrahedra admits 4-simplex states, but no $k$-simplex states for $k \geq 5$.  Going beyond specific examples, for a $d$-dimensional lattice, $k$-simplex states with $k > d + 1$ are impossible, unless the exchange couplings are fine-tuned.  To see this, consider the $k$ points of a simplex in $d$-dimensional space.  Any pair $(\br, \br')$ of these points must have
${\cal J}_{\br \br'} = {\cal J}_{{\rm max}}$; this can only be achieved without fine-tuning if space group symmetry forces all the exchange couplings to be equal.  This can occur only if the points of the simplex are mutually equidistant, and there can be at most $d+1$ mutually equidistant points in $d$-dimensional space.

While on a given lattice there may be other states that saturate the bound even when no $k$-simplex VCS states exist, for large enough $k$ saturation is impossible.  To illustrate this, consider again a lattice where either ${\cal J}_{\br \br'} = {\cal J}_{{\rm max}}$, or ${\cal J}_{\br \br'} = 0$, and let ${\cal N}_b$ be the total number of bonds in the lattice with nonzero exchange.  We can obtain a lower bound on the energy by treating each bond as an isolated system, calculating the resulting two-site ground state energy, and summing over bonds. In Appendix~\ref{app:2site-exact} it is shown that the ground state energy of an isolated bond is $- n_c N {\cal J}_{{\rm max}} / k^2$, so we have
\begin{equation}
\label{eqn:bond-gs-bound}
E_{{\rm MFT}}  \geq - \frac{n_c N {\cal N}_b {\cal J}_{{\rm max}}}{k^2} \text{.}
\end{equation}
This bound is more strict than Eq.~(\ref{eqn:rokhsar-bound}) when $k > 2 {\cal N}_b / N_s + 1$, so saturation of Eq.~(\ref{eqn:rokhsar-bound}) is impossible for such values of $k$.

\section{Large-$N$ limit: Bipartite lattices}
\label{sec:bipartite}

\subsection{Bipartite lower bound}
\label{sec:bipartite-bound}

We now derive a stricter lower bound on the mean-field energy that holds for bipartite lattices.  As in Sec.~\ref{sec:bound}, we consider the mean-field Hamiltonian at general $n_c$, but now on a bipartite lattice.  Precisely, we divide the lattice into two sublattices $A$ and $B$ of equal size so that ${\cal J}_{\br \br'}$ is only nonzero when $\br$ and $\br'$ lie in different sublattices.  We first use the inequality $E_{{\rm MFT}} \geq E'_{{\rm MFT}}$ precisely as in Sec.~\ref{sec:bound}.  The bipartite structure allows us to obtain a stricter bound on $E'_{{\rm MFT}}$.  We recall that
\begin{equation}
{\cal H}'_{{\rm MFT}} = \prsum \frac{N}{{\cal J}_{\br \br'} } \operatorname{tr} ( \bar{\chi}^\dagger_{\br \br'} \bar{\chi}^{\vphantom\dagger}_{\br \br'} )  + {\cal H}_K \text{.}
\end{equation}
The crucial observation is that, for a bipartite lattice, ${\cal H}_{K}$ obeys sublattice symmetry, where ${\cal H}_K \to - {\cal H}_K$ under the operation
\begin{eqnarray}
f_{\br a \alpha} \to \left\{ \begin{array}{ll}
f_{\br a \alpha} & \br \in A \\
- f_{\br a \alpha} & \br \in B \end{array} \right. \text{.}
\end{eqnarray}

Again we let ${\cal L}$ be the set of $n_c m N_s$ occupied levels.  Now, however, we define the set ${\cal U}$ to be the image of ${\cal L}$ under the sublattice operation. 
The set ${\cal U}$ clearly contains only empty levels.  We denote the set of the remaining $n_c (N - 2 m) N_s$ levels by ${\cal M}$.  Levels in ${\cal M}$ are empty and have energies intermediate between those in ${\cal L}$ and ${\cal U}$.  We define averages of $\epsilon_q$ and $\epsilon^2_q$ over these sets as before.

As in Sec.~\ref{sec:bound}, we have $E_K = n_c m N_s [ \epsilon]_{{\cal L}}$, and we need to relate $[\epsilon]_{\cal L}$ to $[\epsilon^2]$.  We have
\begin{eqnarray}
[ \epsilon^2 ] &=& \frac{m}{N} [\epsilon^2]_{\cal L} + \frac{m}{N} [ \epsilon^2]_{\cal U} + \frac{(N - 2 m)}{N} [ \epsilon^2]_{{\cal M}} \\
&=& \frac{2 m}{N} [\epsilon^2]_{\cal L}  + \frac{(N - 2 m)}{N} [ \epsilon^2]_{{\cal M}} \\
&\geq &  \frac{2 m}{N} [\epsilon^2]_{\cal L}  
\geq \frac{2 m}{N} [\epsilon]^2_{\cal L} \text{.}
\end{eqnarray}
Since $[\epsilon]_{\cal L}$ is negative, this implies
\begin{equation}
[\epsilon]_{\cal L} \geq - \sqrt{ \frac{N}{2m} } \sqrt{ [\epsilon^2] } \text{.}
\end{equation}
From this point, we can precisely follow the steps of Sec.~\ref{sec:bound} to minimize the lower bound on $E'_{{\rm MFT}}$.  In this case we obtain the stricter bound
\begin{equation}
\label{eqn:bipartite-bound}
E_{{\rm MFT}} \geq - \frac{1}{4 k} n_c N N_s {\cal J}_{{\rm max}} \text{.}
\end{equation}
This bound is equivalent to Eq.~(\ref{eqn:rokhsar-bound}) when $k = 2$, and is stricter when $k  > 2$.

\subsection{Saturation of the bipartite bound}
\label{sec:bipartite-saturation}

Here we state the necessary and sufficient conditions to saturate the bipartite bound, and give examples of $k$-cluster VCS states that achieve saturation.

The bound Eq.~(\ref{eqn:bipartite-bound}) is saturated if and only if each of the following two conditions hold: 
 (1)  $\epsilon_q$ is constant over each of the sets ${\cal L}$ and ${\cal U}$, and $\epsilon_q = 0$ in ${\cal M}$.  (2) The color density $\tilde{n}^{a b}_{\br}$ calculated using ${\cal H}_K$ satisfies the condition
\begin{equation}
\sum_{\br} \operatorname{tr} (\mu_{\br} \tilde{n}_{\br} ) = 0 \text{.}
\end{equation}
The proof of this statement follows that given for the more general bound in Sec.~\ref{sec:saturation}.  As before, condition (2) comes from saturation of the inequality $E_{{\rm MFT}} \geq E'_{{\rm MFT}}$; since nothing in this inequality depends on the bipartite structure, the proof of condition (2) is identical to that given before.  As before, it is trivial to see that $E'_{{\rm MFT}} = E_{{\rm bound}}$ if and only if condition (1) holds.

As before,  saturation of the bipartite bound is impossible for large enough $k$.  Again we consider a lattice where either ${\cal J}_{\br \br'} = {\cal J}_{{\rm max}}$ or ${\cal J}_{\br \br'} = 0$, and let ${\cal N}_b$ be the total number of bonds in the lattice with nonzero exchange.  For $k > 4 {\cal N}_b / N_s$, the bound Eq.~(\ref{eqn:bond-gs-bound}) is stricter than Eq.~(\ref{eqn:bipartite-bound}), so saturation is impossible for such values of $k$.

\begin{figure}
\includegraphics[width=3in]{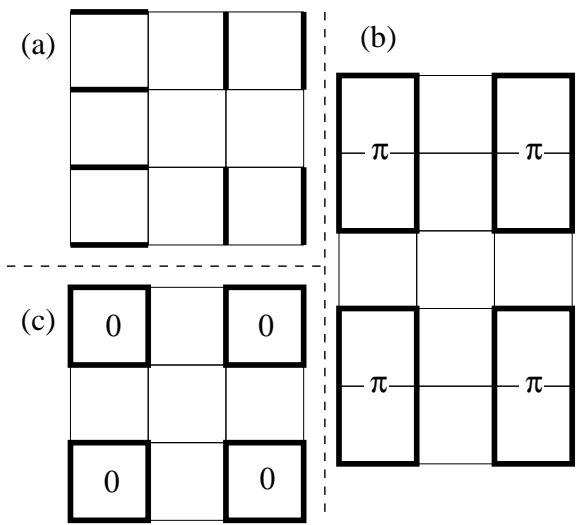}
\caption{Cluster states ($n_c = 1$) with energies saturating the lower bound Eq.~(\ref{eqn:bipartite-bound}) on the square lattice, for $k = 2$ (a), $k = 3$ (b) and $k = 4$ (c). $\chi_{\br \br'}$ has constant magnitude on the dark bonds and is zero on the others.  In the $k = 3$  state, the flux through each six-site plaquette is $\pi$, while it is zero for each four-site plaquette in the $k = 4$ state.  For each value of $k$, in the $N = \infty$ limit, every tiling of the square lattice by the type of clusters shown is a ground state. This large degeneracy is expected to be lifted upon computing perturbative $1/N$ corrections to the ground state energy.\cite{read89b}}
\label{fig:states}
\end{figure}

Since a flat energy spectrum of the mean-field Hamiltonian is necessary to saturate the bipartite bound, we expect that it will only be saturated by VCS states.  VCS states saturating the bound on the square lattice for $n_c = 1$ are shown in Fig.~\ref{fig:states} and were also reported in Ref.~\onlinecite{hermele09}.  For $k=2$ the bound is saturated by any dimer state, and for $k=4$ it is saturated by 4-cluster states of the type shown.  For $k=3$ the bound is actually saturated by a class of 6-cluster states.  

\begin{figure}
\includegraphics[width=3in]{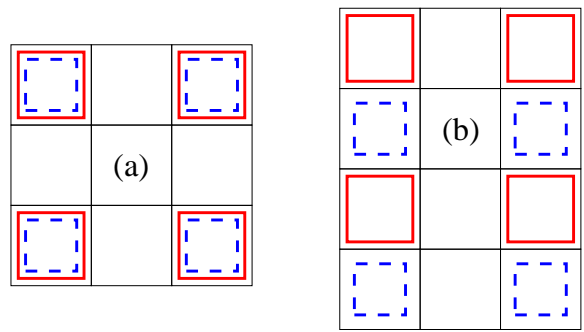}
\caption{Illustration of two $N = \infty$ cluster ground states on the square lattice for $n_c = 2$ and $k = 4$, which saturate the lower bound Eq.~(\ref{eqn:bipartite-bound}).  Square clusters of one color are marked with solid lines (red online), while those of the other color are marked with dashed lines (blue online).  Any configuration where clusters of the two colors separately tile the lattice is a $N = \infty$ ground state -- as in the $n_c = 1$ case, the degeneracy among these states is expected to be lifted upon computing perturbative $1/N$ corrections to the ground state energy.}
\label{fig:nc2-states}
\end{figure}

Whenever a given lattice admits a $n_c = 1$ cluster state saturating the bound, it is easy to see that the same lattice (\emph{i.e.} same set of exchange couplings ${\cal J}_{\br \br'}$) also admits $n_c > 1$ cluster states saturating the bound.  These $n_c > 1$ states have diagonal $\chi^{a b}_{\br \br'}$ as in Eq.~(\ref{eqn:diagonal}), and each $\chi^a_{\br \br'}$ is chosen to give a cluster decomposition of the type that saturates the bound for $n_c = 1$.  Examples of such states (for $k = 4$ and $n_c  =2$) are illustrated for the square lattice in Fig.~\ref{fig:nc2-states}.

\section{Large-$N$ results on square lattice and numerical ground state search}
\label{sec:square}

In this section we focus on the square lattice, and in particular on the case $k \geq 5$.  The discussion of Sec.~\ref{sec:bipartite-saturation} above establishes that, for $k = 2,3,4$, the large-$N$ ground states on the square lattice are VCS states, of the type shown in Figs.~\ref{fig:states} and~\ref{fig:nc2-states}.  
We know of no cluster states that can saturate the bound for $k \geq 5$ on the square lattice, and we conjecture that saturation is impossible for such values of $k$.  In this situation it is very challenging to rigorously determine the large-$N$ ground state, a problem we do not currently know how to solve.  Instead, we employ a systematic numerical search for ground states, which, while not foolproof, allows us to determine the ground state with some confidence.

Here we first describe our numerical self-consistent minimization (SCM) procedure, which we developed and employed in Ref.~\onlinecite{hermele09} for the case $n_c = 1$.  A very similar procedure was later used by M. Foss-Feig and A. M. Rey to study the Kondo lattice model, in collaboration with one of us (M.H.),\cite{fossfeig10a} and subsequently with both of us.\cite{fossfeig10b}  Due to the local constraint, the SCM procedure is not simply a trivial iteration of a self-consistent equation, and to our knowledge it has not been used previously by others;  therefore, we shall describe the SCM procedure here in some detail.  Following this discussion, we shall describe the results of SCM on the square lattice for $n_c =1, 2$.

\subsection{Self-consistent minimization procedure}

We first describe the SCM algorithm in the simpler case of $n_c = 1$; modifications in the $n_c = 2$ case are described below.  The basic idea is simply to iterate the self-consistency condition Eq.~(\ref{eqn:nc1-spe-chi}).  However, if this is all one does, then the fermion density will be non-uniform and Eq.~(\ref{eqn:nc1-spe-constraint}) will be violated.  Instead, the idea is to iterate Eq.~(\ref{eqn:nc1-spe-chi}) within a constrained set of $\chi_{\br \br'}$ and $\mu_{\br}$, so that Eq.~(\ref{eqn:nc1-spe-constraint}) is always satisfied.  To accomplish this, the algorithm proceeds as follows:  (1) An initial $\chi_{\br \br'}$ is chosen randomly.  In our calculations, we chose $\chi_{\br \br'} = |\chi_{\br \br'} | e^{i \phi_{\br \br'}}$, where $|\chi|$ was chosen in the interval [0.03, 0.18] and $\phi$ in the interval $[0, 2\pi]$, both with a uniform distribution.  
(2) Given $\chi_{\br \br'}$, the potential $\mu_{\br}$ is chosen so that Eq.~(\ref{eqn:nc1-spe-constraint}) is satisfied. We describe below how this is done.  (3)  A new set of $\chi$ fields is calculated by
\begin{equation}
\chi'_{\br \br'} = - \frac{{\cal J}_{\br \br'} }{N} \langle f^\dagger_{\br' \alpha} f^{\vphantom\dagger}_{\br \alpha} \rangle \text{.}
\end{equation}
(4) We return to step 2, and iterate until the ground state energy converges.  In practice, we run this procedure for 500 iterations, by which time the convergence is observed to be excellent.

To improve the efficiency of the algorithm as well as its convergence behavior, it is desirable to restrict $\chi_{\br \br'}$ and $\mu_{\br}$ to vary within a unit cell, which is then periodically repeated to form a larger lattice, with periodic boundary conditions.  The translation symmetry generated by the unit cell primitive vectors allows us to exploit Bloch's theorem, further increasing the efficiency.  Since different unit cells can accommodate different candidate ground states, a variety of different cells need to be considered separately.

SCM is indeed a minimization procedure for the ground state energy $E_{{\rm MFT}}$, as the energy is non-increasing for each iteration.  To see this, suppose we have some $\chi_{\br \br'}$ and $\mu_{\br}$ obtained after step 2.   In general this is not a saddle point, but Eq.~(\ref{eqn:nc1-spe-constraint}) is satisfied.  We let $H_{{\rm MFT}}$ be the mean-field Hamiltonian defined in terms of $\chi$ and $\mu$.  We let $\chi'_{\br \br'}$ and $\mu'_{\br}$ be the fields obtained at the next step of the SCM procedure, and $H'_{{\rm MFT}}$ is the mean-field Hamiltonian defined in terms of the primed fields.  We have
\begin{equation}
\chi'_{\br \br'} = - \frac{{\cal J}_{\br \br'}}{N} \langle f^\dagger_{\br' \alpha} f^{\vphantom\dagger}_{\br \alpha} \rangle \text{,}
\end{equation}
where the expectation value $\langle \rangle$ is taken in the ground state of $H_{{\rm MFT}}$.  The potential $\mu'_{\br}$ is chosen so that $\langle f^\dagger_{\br \alpha} f^{\vphantom\dagger}_{\br \alpha} \rangle' = m$, where the primed expectation value is taken in the ground state of $H'_{{\rm MFT}}$.  We have
\begin{equation}
E_{{\rm MFT}} = \langle H_{{\rm MFT}} \rangle = N \prsum \frac{1}{{\cal J}_{\br \br'}} \Big[
| \chi_{\br \br'}|^2 - ( \chi^*_{\br \br'} \chi'_{\br \br'} + \text{c.c.} ) \Big] \text{.}
\end{equation}
Next, we have the variational upper bound
\begin{equation}
E'_{{\rm MFT}} = \langle H'_{{\rm MFT}} \rangle' \leq \langle H'_{{\rm MFT}} \rangle = - N \prsum \frac{ | \chi'_{\br \br'} |^2 }{{\cal J}_{\br \br'} } \text{.}
\end{equation}
Therefore the change in energy satisfies
\begin{equation}
E'_{{\rm MFT}} - E_{{\rm MFT}} \leq -N \prsum \frac{ | \chi_{\br \br'} - \chi'_{\br \br'} |^2 }{ {\cal J}_{\br \br'} } \leq 0 \text{;}
\end{equation}
that is, the energy is non-increasing for every step of the SCM procedure.  This means that when the procedure converges (in practice it almost always does), it converges to a saddle point which is a local minimum of the energy.  There is no guarantee, however, of a global minimum, so, in order to have any confidence that a particular state is the global minimum,  it is necessary to run the procedure many times with different random initial states.

While the other steps of the algorithm are very simple, choosing the potential $\mu_{\br}$ in step 2 requires a more detailed discussion.  The basic idea is to use linear response theory to find the change in potential needed to achieve a desired change in the fermion density.  Going into step 2, we have fields $\chi_{\br \br'}$ and $\mu_{\br}$, which can be used to construct $H_{{\rm MFT}}$.  The density will not in general be uniform, and we define   
\begin{equation}
n_{\br 0} = \langle f^\dagger_{\br \alpha} f^{\vphantom\dagger}_{\br \alpha} \rangle \text{,}
\end{equation}
where the expectation value is taken using the ground state of $H_{{\rm MFT}}$.    Suppose the potential is changed by $\mu_{\br} \to \mu_{\br} + \delta \mu_{\br}$.  To first order in $\delta\mu_{\br}$, the change in the density is
\begin{equation}
\label{eqn:linresp}
\delta n_{\br} = \sum_{\br'} X_{\br \br'} \delta \mu_{\br'} \text{,}
\end{equation}
where $X_{\br \br'} = X_{\br' \br} = X^*_{\br \br'}$ is (by definition) the density response function evaluated in real space and at zero frequency, which, using standard results of linear response theory,  can be calculated from the single-particle wavefunctions and energies of $H_{{\rm MFT}}$.   While straightforward, calculation of $X_{\br \br'}$ is the most computationally expensive step of the algorithm, and must be implemented with attention to efficiency.
At this point, the idea is to set $\delta n_{\br} = m - n_{\br 0}$ (the deviation between the original and desired densities), and invert Eq.~(\ref{eqn:linresp}) to find $\delta\mu_{\br}$.  

In practice $X_{\br \br'}$ is not invertible, because the density does not change under a uniform shift of $\mu_{\br}$ in the canonical ensemble.  Instead we proceed by diagonalizing $X_{\br \br'}$: 
\begin{equation}
\sum_{\br'} X_{\br \br'} u_{\br' \alpha} = x_\alpha u_{\br \alpha} \qquad \text{(no sum on }\alpha\text{).}
\end{equation}
Here $x_{\alpha}$ are the eigenvalues of $X_{\br \br'}$, labeled by $\alpha$, and $u_{\br \alpha}$ are the orthonormal eigenvectors.  If we expand $\delta n_{\br}$ and $\delta\mu_{\br}$ in the basis of eigenvectors, we can rewrite Eq.~(\ref{eqn:linresp}) as
\begin{equation}
\delta n_{\alpha} = x_{\alpha} \delta\mu_{\alpha}  \qquad \text{(no sum on }\alpha\text{).}
\end{equation}
We invert this by simply ignoring eigenvectors with $x_{\alpha} = 0$, and choosing
\begin{equation}
\delta \mu_{\alpha} = \left\{ \begin{array}{ll}
\delta n_{\alpha} / x_{\alpha} & \text{, } x_{\alpha} \neq 0 \\
0 & \text{, } x_{\alpha} = 0 
\end{array} \right. \text{.}
\end{equation}
This is easily converted back to a result for $\delta\mu_{\br}$.

What we have obtained is a linear extrapolation for $\delta\mu_{\br}$, and the basic idea at this point is to proceed by replacing $\mu_{\br} \to \mu_{\br} + \delta \mu_{\br}$, and iterating the procedure until the density is uniform.  This is, in fact, just a multi-dimensional Newton's method for finding a zero of $(n_{\br} - m) = F_{\br}[ \{ \mu_{\br} \}]$.  
While such a method has good local convergence properties (\emph{i.e.} starting sufficiently close to the zero), the global convergence properties are poor.  However,  this can be improved by very simple modifications.\cite{numericalrecipes}  
We define the merit function ${\cal E} = \sum_{\br} (n_{\br} - m)^2$, and demand that each change in $\mu_{\br}$ decrease ${\cal E}$.  If $\mu_{\br} \to \mu_{\br} + \delta \mu_{\br}$ actually increases ${\cal E}$, then we try the smaller step $\mu_{\br} \to \mu_{\br} + \lambda \delta \mu_{\br}$\text{,} where $0 < \lambda < 1$.  This is guaranteed to decrease ${\cal E}$ for sufficiently small $\lambda$; in practice, we use the sequence $\lambda = 1, 0.5, 0.4, 0.3, 0.2, 0.1, 0.09, 0.08, \dots, 0.01, 0.009, \dots$, and give up (simply moving on to step 3) after 1000 attempts.  In practice it is only rarely necessary to give up; even when it is necessary, step 2 is successful in later iterations, and convergence still occurs.  For each iteration of step 2, this process of choosing a new $\delta\mu_{\br}$ by linear extrapolation is repeated 10 times, or until ${\cal E} < 10^{-20}$.  This tolerance for ${\cal E}$ is usually achieved after only a small number of iterations, and is virtually always achieved by the end of a run (500 iterations).

Rarely, it happens that ${\cal E}$ is of order unity after a substantial number of iterations, and the algorithm either converges extremely slowly or fails to converge.  To avoid this problem, when ${\cal E} \geq 1$ any time after 10 iterations, we abort the calculation and start over with a new random initial condition.

We now describe how the SCM procedure is modified to handle $n_c = 2$.  The initial set of $\chi^{a b}_{\br \br'}$ is chosen making use of the singular value decomposition
\begin{equation}
\chi = U \left( \begin{array}{cc} d_1 & 0 \\ 0 & d_2 \end{array}\right) V \text{.}
\end{equation}
Here $d_1$ and $d_2$ are each chosen in the interval $[0, 0.2]$ with a uniform distribution.  $U$ and $V$ are both random ${\rm U}(2)$ matrices, chosen from a uniform distribution on the ${\rm U}(2)$ manifold.  

The algorithm itself proceeds via the same four steps outlined above.  Only in step 2 are the modifications at all nontrivial:  We have to choose $\mu^{a b}_{\br}$ to satisfy $\langle \hat{n}^{a b}_{\br} \rangle = m \delta^{a b}$.  We proceed precisely as above using linear response theory, except that now the necessary linear response equation has a matrix structure and is
\begin{equation}
\label{eqn:nc2-response-eqn}
\delta n^{a b}_{\br} = \sum_{\br'} \sum_{c, d} X^{a b ; c d}_{\br \br'} \delta \mu^{d c}_{\br'} \text{.}
\end{equation}
Since both $\delta n$ and $\delta \mu$ are Hermitian, is is convenient to expand them in a basis of Hermitian matrices labeled by $A, B = 0, \dots, 3$  -- a convenient basis is the  identity matrix ($A=0$) plus the three Pauli matrices ($A = 1,2,3$).  This allows one to recast Eq.~(\ref{eqn:nc2-response-eqn}) in the form
\begin{equation}
\delta n^A_{\br} = \sum_{\br'} \sum_{B} X^{A B}_{\br \br'} \delta \mu^B_{\br} \text{.}
\end{equation}
 Here, it can be shown that $X^{A B}_{\br \br'} = X^{B A}_{\br' \br}$ and is real, so the response function can be diagonalized as described for $n_c = 1$.  Finally, the merit function ${\cal E}$ also needs to be modified, and we choose
 \begin{equation}
 {\cal E} = \sum_{\br} (n^0_{\br} - m)^2 + \sum_{\br} \sum_{A = 1}^3 ( n^A_{\br} )^2 \text{.}
 \end{equation}
 
 \subsection{Results of SCM}
 
We now describe the results of SCM on the square lattice for both $n_c = 1,2$.  The $n_c = 1$ results were reported in Ref~\onlinecite{hermele09}.  Following the protocol described below, we studied $k = 5,6,7,8$ for $n_c = 1$, and $k = 5,6,7$ for $n_c = 2$.  (The numerics become more time-consuming with increasing $k$ and $n_c$.)  We also checked that SCM indeed produces exact ground states (guaranteed by saturation of lower bounds) for smaller values of $k$.

For each value of $k$ and $n_c$ noted above, we considered all unit cells of rectangular geometry containing $k^2$ or fewer lattice sites, excluding cells of unit width for technical reasons.  A unit cell of dimensions $\ell_x \times \ell_y$ is periodically repeated to fill the lattice using Bravais lattice vectors $\bR = \ell_x \bx + \ell_y \by$. (Note that other choices of Bravais lattice vectors are possible -- we made this restriction for the sake of simplicity and limited computation time.)  The lattice itself has periodic boundary conditions and dimensions $L_x \times L_y$.  Letting $L = \operatorname{min} (L_x, L_y)$, we always considered $L \geq 40$ for $k = 5$, $L \geq 36$ for $k = 6$, $L \geq 42$ for $k =7$, and $L \geq 40$ for $k = 8$.  While in some cases we also considered larger system sizes,  a more systematic study of finite-size effects would be desirable, but we have left this for future work.  For each unit cell size, we ran the SCM procedure 30 times, using a different random initial condition each time.  

For $n_c = 1$ and $5 \leq k \leq 8$, we found the ACSL to be the ground state.\cite{hermele09}  For $n_c = 2$ and $k = 5$, we found the ground state to be a rather complicated inhomogeneous state that we have not fully characterized.  On the other hand for $k = 6,7$ we found that the nACSL and dCSL are degenerate ground states.

\section{Discussion}
\label{sec:discussion}

We analyzed a variety of SU$(N)$ symmetric Heisenberg models in two dimensions on the square lattice and gave arguments that topologically ordered spin liquids  are among their ground states. In view of their potential realization with alkaline earth atoms placed on optical lattices, we now summarize what we know about realistically achievable SU$(N)$ Heisenberg models.  Following that discussion, we conclude by mentioning some directions for future study.

The Heisenberg models with $n_c=1$ can be obtained simply as a large-$U$ limit (Mott insulator phase) of a Hubbard model representing alkaline earth atoms hopping on a lattice with $m$ atoms (in their ground electronic state $g$) per site. Such Heisenberg models are within the reach of experiment. \cite{taie10,sugawa10} The main issue is temperature, since the achieved temperature in experiments is in the range $t^2/U < k_B T < U$, and not $k_B T<t^2/U$ ($t$ is the Hubbard hopping) necessary for observing effects of magnetic exchange. Yet this is similar to the issues encountered in studying the SU$(2)$ Hubbard model with cold alkali atoms, and currently a significant amount of effort is being spent trying to devise techniques to lower the temperature of Mott insulators. Assuming this is done, the study of the $n_c=1$ Heisenberg model will be possible in the future.

\begin{figure}
\includegraphics[width=3.5in]{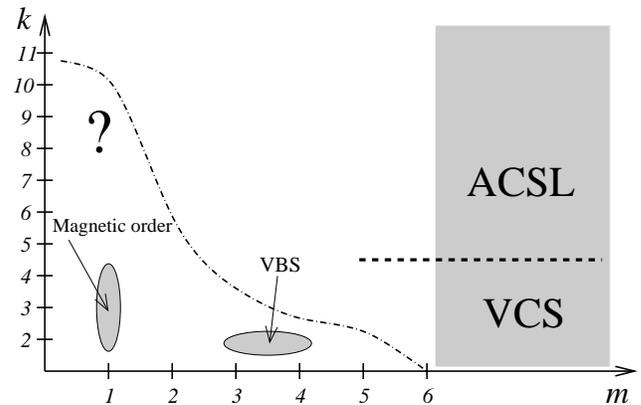}
\caption{Phase diagram of the SU$(N)$ Heisenberg model in two dimensions on the square lattice with $n_c=1$ and with $N=mk$.  In terms of an underlying Hubbard model, $m$ is the number of fermions per site, while $k$ is the inverse filling.  Regions where there is substantial evidence for a given ground state -- or where the ground state is known -- are shaded.  The Abelian chiral spin liquid (ACSL) and valence cluster state (VCS) regions on the right are established by our large-$N$ analysis; the boundary between these regions in large-$N$ is shown by a dashed line.
For $k=2$, $m=1$ the Neel state is the well-known ground state.  There is also evidence for magnetic order at $k=3$, $m=1$\cite{toth10} and $k=4$, $m=1$.\cite{corboz11} Valence-bond solid (VBS) order (which is a type of VCS) was found for $k=2$ and $m=3,4$.\cite{assaad05} The dashed-dot line separates the range of parameters beyond the reach of current experiments (above and to the right of the line) and the range within the reach of the experiments (below and to the left of the line).  The experimentally relevant part of the phase diagram with the greatest potential for novel ground states -- in particular, the Abelian chiral spin liquid -- is indicated with a question mark.} 
\label{fig:phasediagram}
\end{figure}

We summarize what we know about the $n_c=1$ Heisenberg model in Fig.~\ref{fig:phasediagram}. On the horizontal axis of this figure, we plot $m$, the number of atoms in
the same electronic state $g$ per site. On the vertical axis we plot $k$, which is $k=N/m$. The dashed-dotted line represents roughly the curve $km=10$. The significance
of this curve lies in the fact that $km=N$ and $N=10$ is the largest experimentally achievable $N$. Therefore, all the points on the plot which lie above the curve $km=10$ cannot
be reached experimentally while those below the curve can. The actual curve on Fig.~\ref{fig:phasediagram} is corrected to take into account that $k$ and $m$ are integers.

We emphasize that any $N \le 10$ is within reach of an experiment. Indeed, working with $^{87}$Sr, for example, one can selectively populate its nuclear spin states so that only a subset of those are populated with a total number of populated states equal to $N$.\cite{gorshkov10} At the same time, we expect that $m=1$ and $m=2$ columns of the figure are easiest to reach, as higher $m$ will likely experience losses due to 3-body recombination. 

At $m=1$ and $k=2$, the ground state is of course the N\'eel state.  There is also evidence for magnetic order at $m=1$, $k=3$\cite{toth10}  and at $m=1$ and $k=4$.\cite{corboz11}  For $k=2$ and $m \geq 3$, 
it is believed that the ground state is a valence bond solid. This is established by quantum Monte Carlo for $m = 3,4$,\cite{assaad05} and is proven in the limit $m \to \infty$.\cite{rokhsar90}

In addition to that, in this paper we proved that at $m\rightarrow \infty$, $k<5$, the ground states are valence cluster states, of which valence bond solid is a particular example.
Finally, we have shown that at $k>5$ and at least for $k \le 8$, and possibly for $k>8$ as well, and at $m \rightarrow \infty$, the ground state is the
Abelian chiral spin liquid.  The rest of the phase diagram remains to be filled in.  Of course other phases not discussed here may well be present, and there is some evidence this is the case, in particular at $k=2$, $m=2$.\cite{assaad05}

The experiments will be conducted at $m=1$ or $m=2$, and at $k$ as large as 10. The ground state of the Heisenberg model under these conditions is not known; this is represented by a question mark in Fig.~\ref{fig:phasediagram}. We believe  it is unlikely that the N\'eel state can survive to large $k$, even at $m=1$. Indeed, as discussed earlier, the amount of frustration increases with increasing $k$.\cite{hermele09}
What happens in this region needs to be investigated further.  Unfortunately, numerical study is difficult, especially since these models [except when $k=2$ (Ref.~\onlinecite{assaad05})] suffer from the quantum Monte Carlo minus sign problem, even on bipartite lattices, in both world-line and fermion determinantal approaches.  However, it may be possible to obtain useful information from analytical and density matrix renormalization group studies of quasi-one-dimensional systems.  
Ultimately, experiment will need to tell us what happens in this part of the phase diagram. An intriguing  possibility is that the phase boundary which lies between $k=4$ and $k=5$ extends all the way from large $m$ to $m=1$, thus making the experimentally accessible $m=1$, $k>4$ regime a chiral spin liquid. 

We note that, while we only considered integer $k$, some non-integer values of $k$ are possible.  For example, $m = 2$ and $N = 5$ corresponds to $k = 5/2$, and a well-defined large-$N$ limit with $k = 5/2$ certainly exists.  We did not consider such values of $k$ first for simplicity, and second because non-integer $k$ requires $m \geq 2$, making experimental accessibility somewhat less favorable.  Nonetheless, it would be interesting to study the large-$N$ limit for non-integer values of $k$ in future work.

A similar phase diagram can be discussed at $n_c=2$ where the Abelian chiral spin liquid will be replaced by the  non-Abelian chiral spin liquid (or by the doubled chiral spin liquid). 

Supposing that some of the topological liquids discussed here do indeed occur for physically realizable ${\rm SU}(N)$ spin models, it will be an interesting question how to actually observe fractional or non-Abelian statistics in these systems.  This is especially so given the intense interest in topological quantum computation using non-Abelian particles.  We expect that holes, the excitations obtained by removing an atom from the system,
 should split into spinons and holons. The holons may be localized
near a given site by an external potential and, at the same time, they obey fractional or non-Abelian statistics depending on which topological liquid we are considering.  Therefore braiding may be achieved by manipulating the holons via the external potential, and this is a route by which fractional and non-Abelian statistics may be observed. While some further details along these lines are given in Appendix~\ref{app:carriers}, many questions remain open, and we feel this constitutes an interesting direction for future work.

Other directions for future study include investigation of the projected wavefunctions for the various topological liquids, which we discussed only briefly in Sec.~\ref{sec:top}.  Given the difficulty of unbiased numerical study in these systems, such wavefunctions may be useful as variational states to gain understanding of the phase diagram away from the large-$N$ limit.  Finally, another potentially interesting problem is a careful study of the dCSL edge states, which may be topologically protected as mentioned in Sec.~\ref{sec:dcsl}.

\begin{acknowledgments}
We are grateful to Gang Chen, Charles Kane,  Andreas La\"{u}chli, Hao Song and Ashvin Vishwanath for useful discussions, and are especially grateful to Ana Maria Rey both for numerous useful discussions and ongoing related collaborations.  This research is supported by DOE award no. DE-SC0003910 (M.H.), and NSF grants no. DMR-0449521 (V.G.) and PHY-0904017 (V.G.).

\end{acknowledgments}

\appendix

\section{Alkaline earth atom Hubbard and spin models}
\label{sec:atomic}

Here we briefly review the Hubbard model describing fermionic alkaline earth atoms in optical lattices.  We focus on two kinds of Mott insulating states, in which the spin models we study are the simplest description capturing the essential physics.  A more extensive and detailed discussion of fermionic AEA in optical lattices, and the rich variety of strong correlation physics that can be realized in these systems, can be found in Ref.~\onlinecite{gorshkov10}.

A single alkaline earth atom has a $^{1}S_0$ electronic ground state.  (Recall that the subscript on the right is $J$, the electronic angular momentum, so this state has $J = 0$.)  The nuclear spin can be as large as $I = 9/2$ in the case of $^{87}$Sr.  Other important examples are $^{171}$Yb and $^{173}$Yb, with $I = 1/2$ and $I = 5/2$, respectively. While Yb is not an alkaline earth, it has the same configuration of outer electrons, and all the discussion here applies equally to alkaline earths and to Yb.  Also important for our purposes is the $^{3}P_0$ lowest electronic excited state, which has a very long lifetime on the order of $100\, {\rm s}$.  These two electronic states can be subjected to optical lattices of different strength.\cite{daley08}

Interactions between two atoms in any combination of these electronic states, which arise from collisions in the $s$-wave channel, are expected to respect a large ${\rm SU}(N)$ spin rotation symmetry, where $N = 2 I + 1$ is the number of nuclear spin levels per atom.\cite{gorshkov10, cazalilla09}  The symmetry arises because such atoms have $J = 0$, and due to the resulting quenching of hyperfine coupling, the nuclear spin is essentially a spectator in the collision between two atoms, and only participates via Fermi statistics.  The ${\rm SU}(N)$ symmetry is not exact but is expected to hold to an excellent approximation.  A rough estimate is that, for two ground state atoms, ${\rm SU}(N)$-breaking effects are $10^{-9}$ times the strength of the ${\rm SU}(N)$-symmetric interaction.\cite{gorshkov10}  For two excited state atoms, the strength of ${\rm SU}(N)$-breaking is estimated to be $10^{-3}$.

We now suppose that the atoms are subjected to an optical lattice potential deep enough that a description in terms of a one-band Hubbard model is appropriate.  We introduce creation operators $c^\dagger_{\br g \alpha}$ and $c^\dagger_{\br e \alpha}$ for the ground state ($g$) and excited state ($e$) atoms, respectively.  Here, $\br$ labels the lattice site, and $\alpha = 1,\dots,N$ labels the $z$-component of nuclear spin. (We shall find this notation more convenient than $I_z = -I, \dots, I$, due to the ${\rm SU}(N)$ symmetry.)  To describe the system we consider the most general Hubbard model with ${\rm SU}(N)$ symmetry, nearest-neighbor hopping, and on-site interactions.  It is also important to note that the numbers of ground state and excited state fermions are separately conserved, due to the long lifetime (treated here as infinite) of the excited state fermions, and energy conservation.
  The Hamiltonian is\cite{gorshkov10}
\begin{eqnarray} \label{eq:gorshkov}
H &=& -t_g \sum_{\langle \br \br' \rangle} ( c^\dagger_{\br g \alpha} c^{\vphantom\dagger}_{\br' g \alpha} + \text{H.c.} ) 
- t_e \sum_{\langle \br \br' \rangle} ( c^\dagger_{\br e \alpha} c^{\vphantom\dagger}_{\br' e \alpha} + \text{H.c.} ) \nonumber \\
&+& \sum_{\br} \big( \frac{U_{gg}}{2} n^2_{\br g} +  \frac{U_{e e}}{2} n^2_{\br e} + U_{e g}  n_{\br g} n_{\br e}  \big) \nonumber \\
&-& J_{eg} \sum_{\br} S^g_{\alpha \beta}(\br) S^e_{\beta \alpha} (\br) \text{.} \label{eqn:hubb}
\end{eqnarray}
The sums in the first two terms are over nearest-neighbor bonds.  We have introduced the following number and spin operators:
\begin{eqnarray}
n_{\br g} &=& \sum_{\alpha} c^\dagger_{\br g \alpha} c^{\vphantom\dagger}_{\br g \alpha} \\
S^g_{\alpha \beta}(\br) &=& c^\dagger_{\br g \alpha} c^{\vphantom\dagger}_{\br g \beta} \text{,}
\end{eqnarray}
with corresponding expressions for $n_{\br e}$ and $S^e_{\alpha \beta}(\br)$.
The on-site interaction parameters $U_{gg}$, $U_{ee}$, $U_{eg}$ and $J_{eg}$ are proportional to linear combinations of the four independent $s$-wave scattering lengths characterizing collisions among the atoms.\cite{gorshkov10}

The ${\rm SU}(N)$ spin symmetry acts on the fermions as follows:
\begin{eqnarray}
c^\dagger_{\br g \alpha} &\to& U_{\alpha \beta} c^\dagger_{\br g \beta} \nonumber \\
c^\dagger_{\br e \alpha} &\to& U_{\alpha \beta} c^\dagger_{\br e \beta} \text{.}
\end{eqnarray}
Here, $U$ is an arbitrary ${\rm SU}(N)$ matrix.  The fermions thus transform in the fundamental representation of ${\rm SU}(N)$.  

We shall consider $U_{gg} > 0$, which is known to be the case for $^{87}$Sr and $^{173}$Yb.  In both cases the corresponding scattering length is about $100 \, a_0$, which corresponds to rather large repulsive interactions.\cite{enomoto08,escobar08}  The sign of the interspecies exchange interaction $J_{e g}$ is not yet known and may be either ferromagnetic (positive) or antiferromagnetic (negative); this is likely to depend on the atomic species.  If one ground state atom and one excited state atom share the same site, antiferromagnetic (ferromagnetic) $J_{eg}$ favors antisymmetric (symmetric) combinations of their nuclear spins.

We consider two types of Mott insulators.  The simpler of the two is realized using only ground state atoms, at an integer filling of $m$ atoms per site.  While $m=1$ best avoids issues of three-body loss, we consider general $m$ because it is needed for the large-$N$ limit.  In this case, the Hubbard model contains only the $t_g$ and $U_{gg}$ terms, and when $t_{g} \ll U_{gg}$ the standard degenerate perturbation theory\cite{auerbachbook} gives the spin model
\begin{equation}
\label{eqn:hspin}
H_{{\rm spin}} = J \sum_{\langle \br \br' \rangle} S_{\alpha \beta}(\br) S_{\beta \alpha}(\br') \text{,}
\end{equation}
where $J = 2 t_g^2 / U_{gg}$, and we have defined
\begin{equation}
S_{\alpha \beta}(\br) = S^g_{\alpha \beta}(\br) + S^e_{\alpha \beta}(\br) \text{.}
\end{equation}
In this case, since no excited state atoms are present, $S_{\alpha \beta}(\br) = S^g_{\alpha \beta}(\br)$. The spin at each site transforms in the $m \times 1$ irreducible representation of  ${\rm SU}(N)$ (Fig.~\ref{fig:rect_youngtab}); this simply expresses the fact that the nuclear spins of the identical fermions are combined antisymmetrically.  For $m = 1$, the spin transforms in the fundamental representation of ${\rm SU}(N)$, and when $N=2$ this is simply a $S = 1/2$ spin.

The second type of Mott insulator is related to $S = 1$ Mott insulators of ${\rm SU}(2)$ spins.  It is realized with $m$ ground state atoms \emph{and} $m$ excited state atoms on each site.  
We consider $J_{eg} > 0$, in which case the single-site ground state is associated with a $m \times 2$ tableau. This can be seen by first viewing the ground state atoms as forming a spin transforming in the $m \times 1$ representation, which is required simply by Fermi statistics, and similarly for the excited state atoms.  These two spins are then coupled by the $J_{eg}$ exchange term, and formally we need to solve a two-site problem, which is done in Appendix~\ref{app:2site-exact}.  We do not consider antiferromagnetic interspecies exchange, because this gives a single-site ground state with a $2m \times 1$ tableau, which is of the same type obtained with only ground state atoms

In the simple case $m=1$, the single-site ground states  are of the form
\begin{equation}
| \psi_{\br} \rangle = (c^\dagger_{\br g \alpha} c^\dagger_{\br e \beta} + c^\dagger_{\br g \beta} c^\dagger_{\br e \alpha} ) | 0 \rangle \text{;} \label{eqn:ge-single-site}
\end{equation}
 that is, the nuclear spins of the two fermions are combined symmetrically. When $N = 2$ and $m=1$, this is simply a $S = 1$ spin.  More generally the single-site ground states can be obtained from the  highest-weight state
 \begin{equation}
| \psi^{{\rm hw}}_{\br} \rangle = c^\dagger_{\br g 1} c^\dagger_{\br e 1} \cdots  c^\dagger_{\br g m} c^\dagger_{\br e m}  | 0 \rangle \text{,}
\end{equation}
where all other single-site ground states can be obtained by repeated action on $| \psi^{{\rm hw}}_{\br} \rangle$ with appropriate components of $S_{\alpha \beta}(\br)$.  That is, they are linear combinations of states of the form $S_{\alpha \beta}(\br) | \psi^{{\rm hw}}_{\br} \rangle$, $S_{\alpha \beta}(\br) S_{\gamma \delta}(\br) | \psi^{{\rm hw}}_{\br} \rangle$, and so on.

   Again, $m=1$ best avoids issues of three-body loss, but we shall consider general $m$.  Another potentially important loss mechanism is inelastic losses in collisions between two excited state atoms.  This can be minimized by making the lattice for the excited state atoms very deep, effectively setting $t_e = 0$.
 
In Sec.~\ref{sec:models}, the type of ${\rm SU}(N)$ spin is specified by the two local constraint equations Eq.~(\ref{eqn:num-constraint}) and Eq.~(\ref{eqn:color-constraint}), and in Appendix~\ref{app:1site-irrep} it is shown that these two constraints imply that the spin transforms in the $m \times n_c$ representation.  To make contact with that discussion, we now show that single-site ground states of the present Hubbard model, transforming in the $m \times 2$ representation, satisfy the constraint Eq.~(\ref{eqn:color-constraint}), that is
\begin{equation}
\label{eqn:appendix-color-constraint}
T^i_{\br}  | \psi_{\br} \rangle = 0\text{,}
 \end{equation}
 where
 \begin{equation}
 T^i_{\br}  = \frac{1}{2}  c^\dagger_{\br a \alpha} \sigma^i_{a b} c^{\vphantom\dagger}_{\br b \alpha}  \text{.}
 \end{equation}
 Here, $a,b = e,g$, and we formally consider the $e,g$ labels as an index transforming in an ${\rm SU}(2)$ ``orbital'' space. 
Moreover, $\sigma^i$ are the $2 \times 2$ Pauli matrices ($i = 1,2,3$), and $| \psi_{\br} \rangle$ is a single-site ground state for the site $\br$.  [Since there are $2 m$ fermions on each site, the constraint Eq.~(\ref{eqn:num-constraint}) is obviously satisfied.]

The constraint Eq.~(\ref{eqn:appendix-color-constraint}) is obviously satisfied for the $m =1$ state given in Eq.~(\ref{eqn:ge-single-site}); the wavefunction is antisymmetric under interchange $e \leftrightarrow g$ and is thus an orbital singlet.  The same holds for the highest-weight state $|\psi^{{\rm hw}}_{\br} \rangle$, since it is built as a product of orbital singlets $c^\dagger_{\br g \alpha} c^\dagger_{\br e \alpha}$ (no sum on $\alpha$).   Because $[ S_{\alpha \beta}(\br) , T^i_{\br} ] = 0$, this immediately implies that Eq.~(\ref{eqn:appendix-color-constraint})  holds for all single-site ground states.

When $t_e = 0$ and $t_g \ll U_{gg} , J_{eg}$, the spin Hamiltonian is given by the same form Eq.~(\ref{eqn:hspin}), only now $J = t_g^2/ [ 2 (U_{g g} + J_{eg}) ]$.  The degenerate perturbation theory calculation needed to establish this, unlike in the case of only ground state atoms, is not simply a trivial generalization of the familiar calculation for the $S = 1/2$, ${\rm SU}(2)$ Hubbard model.  While the end result of this calculation appeared in Ref.~\onlinecite{gorshkov10}, the details were not presented, so we now present them here.
 
 First we consider a single lattice site, and note that the energy of $|\psi^{{\rm hw}}_{\br} \rangle$ (neglecting hopping) is
 \begin{equation}
 E_0 = \frac{1}{2} U_{gg} m^2 + \frac{1}{2} U_{ee} m^2 + U_{eg} m^2 - J_{eg} m \text{.}
 \end{equation}
 By ${\rm SU}(N)$ symmetry, this holds for any single-site ground state.  Moreover, we note that
 \begin{equation}
 \label{eqn:dpt-identity}
 c^\dagger_{\br g \alpha} c^{\vphantom\dagger}_{\br e \alpha} | \psi^{{\rm hw}}_{\br} \rangle
 =  c^\dagger_{\br e \alpha} c^{\vphantom\dagger}_{\br g \alpha} | \psi^{{\rm hw}}_{\br} \rangle
 = 0 \text{,}
 \end{equation}
 which also holds for any single-site ground state by ${\rm SU}(N)$ symmetry.
 
Now we consider second-order degenerate perturbation theory for a two-site problem with adjacent lattice sites $\br_1$ and $\br_2$.  (In second-order perturbation theory, there is no need to consider more than two sites.)  We construct the effective Hamiltonian by building up its action on an arbitrary state $| \psi^1_{\br_1} \rangle  | \psi^2_{\br_2} \rangle$ in the low-energy manifold (that is, $| \psi^1_{\br_1} \rangle$ and $| \psi^2_{\br_2} \rangle$ are arbitrary single-site ground states).  The energy of the initial state is $2 E_0$.  The intermediate state is obtained by hopping a single ground state fermion from $\br_1$ to $\br_2$ or vice versa.  We consider hopping from $\br_1$ to $\br_2$ so the intermediate state is
\begin{equation}
| \psi_{{\rm int}} \rangle =  \sum_{\alpha} | \phi^1_{\br_1 \alpha} \rangle | \phi^2_{\br_2 \alpha} \rangle \text{,}
\end{equation}
where
\begin{eqnarray}
| \phi^1_{\br_1 \alpha} \rangle &=& c^{\vphantom\dagger}_{\br_1 g \alpha} | \psi^1_{\br_1} \rangle \\
| \phi^2_{\br_2 \alpha} \rangle &=& c^\dagger_{\br_2 g \alpha} | \psi^2_{\br_2} \rangle \text{.}
\end{eqnarray}
Acting on the intermediate state with the on-site part of the Hamiltonian and using the identity Eq.~(\ref{eqn:dpt-identity}) to evaluate the action of the $J_{eg}$ exchange term, the energy of the intermediate state is found to be
\begin{equation}
E_{{\rm int}} = 2 E_0 + U_{gg} + J_{eg} \text{.}
\end{equation}

Since the intermediate state is an eigenstate with energy independent of the initial state, the effective Hamiltonian is
\begin{eqnarray}
H_{{\rm eff}} &=& \frac{ - t^2_g }{ U_{gg} + J_{eg} } \Big[ {\cal P} c^\dagger_{\br_1 g \alpha} c^{\vphantom\dagger}_{\br_2 g \alpha} (1 - {\cal P} ) c^\dagger_{\br_2 g \beta} c^{\vphantom\dagger}_{\br_1 g \beta} {\cal P} \nonumber \\ &+& (\br_1 \leftrightarrow \br_2)  \Big] \text{,}
\end{eqnarray}
where ${\cal P}$ is the usual projector onto the ground state manifold, and the second term in the square brackets accounts for the process where a fermion first hops from $\br_2$ to $\br_1$.  Because a single hopping process always leaves the ground state manifold (since it changes the fermion number on each site), we can drop the $(1 - {\cal P})$ factor and write
\begin{eqnarray}
H_{{\rm eff}} &=& \frac{ - t^2_g }{ U_{gg} + J_{eg} } \Big[ {\cal P} c^\dagger_{\br_1 g \alpha} c^{\vphantom\dagger}_{\br_2 g \alpha}c^\dagger_{\br_2 g \beta} c^{\vphantom\dagger}_{\br_1 g \beta} {\cal P}+ (\br_1 \leftrightarrow \br_2)  \Big] \nonumber \\
&=&
\frac{ 2 t^2_g}{U_{gg} + J_{eg} } \Big[ {\cal P} S^g_{\alpha \beta}(\br_1) S^g_{\beta \alpha} (\br_2) {\cal P} \Big] \text{,}
\end{eqnarray}
where we dropped an additive constant in going to the second line.  Now,
\begin{equation}
S^g_{\alpha \beta}(\br) = \frac{1}{2} S_{\alpha \beta}(\br) + \frac{1}{2} \big[ S^g_{\alpha \beta}(\br) - S^e_{\alpha \beta}(\br) \big] \text{,}
\end{equation}
where the second term transforms as a triplet in the orbital space.  But the projector ${\cal P}$ forces every lattice site to be an orbital singlet, and therefore
\begin{equation}
{\cal P} S^g_{\alpha \beta}(\br_1) S^g_{\beta \alpha} (\br_2) {\cal P} = \frac{1}{4} S_{\alpha \beta}(\br_1) S_{\beta \alpha}(\br_2) \text{,}
\end{equation}
so that
\begin{equation}
H_{{\rm eff}} = \frac{ t^2_g}{2 (U_{gg} + J_{eg} )} S_{\alpha \beta}(\br_1) S_{\beta \alpha}(\br_2) \text{,}
\end{equation}
the result we claimed above.

\section{Determining the irreducible representation of ${\rm SU}(N)$ spin from local constraints}
\label{app:1site-irrep}

Here, we show that the local constraints Eqs.~(\ref{eqn:num-constraint},\ref{eqn:color-constraint}) imply that each spin  transforms in the $m \times n_c$ irreducible representation of ${\rm SU}(N)$.  Another way to state this fact is that the Hilbert space of a single lattice site, subject to the local constraints, transforms irreducibly under ${\rm SU}(N)$ in the $m \times n_c$ representation.

To see this, it is helpful to think of spin and color rotations as a subgroup ${\rm SU}(N) \times {\rm SU}(n_c) \subset {\rm SU}(n_c N)$, where the fermions transform in the fundamental of ${\rm SU}(n_c N)$.  By fermion antisymmetry, the first constraint [Eq.~(\ref{eqn:num-constraint})] implies that each site transforms in the $n_c m \times 1$ representation of ${\rm SU}(n_c N)$.  

\begin{figure}
\includegraphics[width=2in]{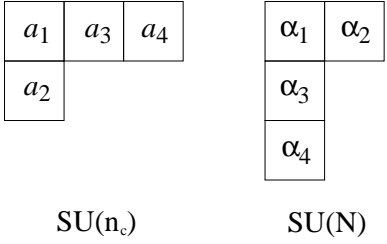}
\caption{Illustration of Young tableaux occurring in the decomposition of the $n_c m \times 1$ representation of ${\rm SU}(n_c N)$ into irreducible representations of the ${\rm SU}(N) \times {\rm SU}(n_c)$ subgroup, for the case $n_c = m = 2$.  If, as described in the text, we project out the ${\rm SU}(n_c)$ irreducible representation corresponding to the tableau on the left, then the corresponding ${\rm SU}(N)$ tableau is as shown on the right.  Note that the rows (columns) of the ${\rm SU}(n_c)$ tableau become the columns (rows) of the ${\rm SU}(N)$ tableau.}
\label{fig:youngexample}
\end{figure}

To understand the role of the second constraint [Eq.~(\ref{eqn:color-constraint})], we need to understand how this representation decomposes into irreducible representations of ${\rm SU}(N) \times {\rm SU}(n_c)$.  The decomposition has the general form
\begin{equation}
\label{eqn:decomp}
( n_c m \times 1)_{{\rm SU}(n_c N) } = \sum_i r^i_{{\rm SU}(N) } \otimes r^i_{{\rm SU}(n_c) }
\end{equation}
This equation expresses the fact that the $(n_c m \times 1)$ representation of ${\rm SU}(n_c N)$ is a direct sum of irreducible representations of ${\rm SU}(N) \times {\rm SU}(n_c)$, labeled by $i$.  We will first show that, for each term in this decomposition, $r^i_{{\rm SU}(N)}$ \emph{uniquely} determines $r^i_{{\rm SU}(n_c)}$, and vice versa.

Focusing on a single lattice site and dropping the site label for fermion operators, we consider the following (overcomplete) basis states for the $( n_c m \times 1)_{{\rm SU}(n_c N) }$ representation:
\begin{equation}
|  a_1, \alpha_1 ;   \dots ; a_{n_c m}, \alpha_{n_c m} \rangle \equiv
 f^\dagger_{a_1 \alpha_1} \dots f^\dagger_{a_{n_c m} \alpha_{n_c m}} | 0 \rangle \text{.}
 \end{equation}
If $P(i)$ is a permutation of the integers $i = 1,\dots,n_c m$, then fermion antisymmetry implies
\begin{eqnarray}
&& \qquad |  a_{P(1)}, \alpha_{P(1)} ;   \dots ; a_{P(n_c m)}, \alpha_{P(n_c m)} \rangle = \nonumber \\
&&  \operatorname{sgn} P |  a_1, \alpha_1 ;   \dots ; a_{n_c m}, \alpha_{n_c m} \rangle \text{,} \label{eqn:antisymmetry}
\end{eqnarray}
where $\operatorname{sgn} P$ is the sign of the permutation.  Suppose we want to project out a particular representation of ${\rm SU}(n_c)$.  We do this by forming a ${\rm SU}(n_c)$ Young tableau with $n_c m$ boxes, and associating each box with a color index $a_i$.  We then follow the usual procedure of first antisymmetrizing the $a_i$ indices occupying the same column, and second symmetrizing those occupying the same row.  Because of the overall antisymmetry expressed in Eq.~(\ref{eqn:antisymmetry}), when in the first step we antisymmetrize the $a_i$ indices in a given column, we also simultaneously \emph{symmetrize} the corresponding set of $\alpha_i$ indices.  Similarly, the second step antisymmetrizes those $\alpha_i$ indices corresponding to a given row.  This means that, in the process of projecting out a given desired ${\rm SU}(n_c)$ representation, we have also automatically projected out a corresponding given ${\rm SU}(N)$ representation.  The tableau of the ${\rm SU}(N)$ representation is given by interchanging the role of rows and columns of the ${\rm SU}(n_c)$ tableau -- see Fig.~\ref{fig:youngexample} for an example that clarifies the meaning of this statement.

The constraint Eq.~(\ref{eqn:color-constraint}) dictates that we keep only the terms in the decomposition where $r^i_{{\rm SU}(n_c)}$ is the singlet representation $0_{{\rm SU}(n_c)}$.  Since we have to form the corresponding tableau using $n_c m$ boxes, the only possible ${\rm SU}(n_c)$ tableau is $n_c \times m$, and the above discussion implies that the corresponding ${\rm SU}(N)$ tableau is $m \times n_c$.  It can be seen by directly constructing a highest weight state that the representation $(m \times n_c)_{{\rm SU}(N) } \otimes 0_{{\rm SU}(n_c)}$ only occurs once in the decomposition.  Therefore the constraint gives
\begin{equation}
( n_c m \times 1)_{{\rm SU}(n_c N) } \to   (m \times n_c)_{{\rm SU}(N)} \otimes 0_{{\rm SU}(n_c) } \text{,}
\end{equation}
the desired result.

\section{Exact ground state energy of two-site problem}
\label{app:2site-exact}

Here we consider a problem of two spins at $\br_1$ and $\br_2$, coupled by the Hamiltonian Eq.~(\ref{eqn:hspin2}). 
We write $J_{\br_1 \br_2} = {\cal J} / N$, so that
\begin{equation}
{\cal H} = \frac{{\cal J}}{N} S_{\alpha \beta}(\br_1) S_{\beta \alpha}(\br_2) \text{.}
\end{equation}
We shall calculate the exact (\emph{i.e.} not large-$N$) ground state energy for arbitrary $N$, $m = N/k$ and $n_c$. 

It is convenient to define the Hermitian spin operators
\begin{equation}
\hat{T}^{{\cal A}}_{\br} = f^\dagger_{\br a \alpha} T^{\cal A}_{\alpha \beta} f^{\vphantom\dagger}_{\br a \beta} \text{,}
\end{equation}
where ${\cal A} = 1, \dots, N^2 - 1$ labels the ${\rm SU}(N)$ generators $T^{{\cal A}}$.  These are chosen to satisfy the orthonormality condition
\begin{equation}
\operatorname{tr} ( T^{\cal A} T^{\cal B} ) = \frac{1}{2} \delta^{ {\cal A} {\cal B} } \text{,}
\end{equation}
and can be shown to satisfy the identity
\begin{equation}
T^{\cal A}_{\alpha \beta} T^{\cal B}_{\gamma \delta} = \frac{1}{2} \Big( \delta_{\alpha \delta} \delta_{\beta \gamma} - \frac{1}{N} \delta_{\alpha \beta} \delta_{\gamma \delta} \Big) \text{.} \label{eqn:sun-gen-identity}
\end{equation}
Equation~(\ref{eqn:sun-gen-identity}) can be used to show
\begin{eqnarray}
{\cal H} &=& \frac{2 {\cal J}}{N} \hat{T}^{\cal A}_{\br_1} \hat{T}^{\cal A}_{\br_2} + \frac{ {\cal J} n_c^2 m^2}{N^2} \\
&=& \frac{ {\cal J}}{N} \Big[  ( \hat{T}^{\cal A}_{\br_1} + \hat{T}^{\cal A}_{\br_2} )^2 - ( \hat{T}^{\cal A}_{\br_1} )^2
-  (\hat{T}^{\cal A}_{\br_2})^2  \Big] + \frac{ {\cal J} n_c^2 m^2}{N^2} \text{.} \nonumber
\end{eqnarray}

Now, $(\hat{T}^{\cal A})^2 = \hat{T}^{\cal A} \hat{T}^{\cal A}$ is the quadratic Casimir of ${\rm SU}(N)$.  In a given irreducible representation $r$ this operator is proportional to the identity, and its eigenvalue $C_2(r)$ can be computed from the structure of the Young tableau using a formula given in Ch.~19 of Ref.~\onlinecite{fsbook}, which we now reproduce.  
Suppose the Young tableau has $n_{row}$ rows, each with length $b_i$ ($i = 1,\dots,n_{row}$) and $n_{col}$ columns, each with length $a_i$ ($i = 1, \dots, n_{col}$), and a total of $\ell$ boxes.  Then the eigenvalue of the Casimir is given by
\begin{equation}
\label{eqn:casimir}
C_2(r) = \frac{1}{2} \Big[ \ell ( N - \ell / N) + \sum_{i = 1}^{n_{row}}  b_i^2 - \sum_{i = 1}^{n_{col}}  a_i^2 \Big] \text{.}
\end{equation}

Since each spin transforms in the $m \times n_c$ representation, we can use Eq.~(\ref{eqn:casimir}) to evaluate $ ( \hat{T}^{\cal A}_{\br_1} )^2 = (\hat{T}^{\cal A}_{\br_2})^2$.  Moreover, by examining the Young tableaux appearing in the tensor product $(m \times n_c) \otimes (m \times n_c)$, and using Eq.~(\ref{eqn:casimir}) to evaluate $ ( \hat{T}^{\cal A}_{\br_1} + \hat{T}^{\cal A}_{\br_2} )^2$ for each tableau, we find that the two-spin ground state is the $2m \times n_c$ tableau, and that the corresponding ground state energy is
\begin{equation}
E_0 =  - \frac{ {\cal J} n_c m^2}{N} = - \frac{ n_c N {\cal J}}{k^2} \text{.}
\end{equation}

\section{Energy of $k$-cluster states}
\label{app:kcluster}

Here we compute the large-$N$ ground state energy of a single isolated $k$-cluster, a result which is used in the discussion of Sec.~\ref{sec:simplex}.  We consider a spin model defined on an arbitrary connected graph with $k$ sites labeled by $s$, and with links labeled by $\ell$.  The exchange energy is taken to be equal on all links and is $J = {\cal J} / N$.  The mean-field Hamiltonian is
\begin{equation}
H_{{\rm MFT}} = \frac{N}{\cal J} \sum_{\ell} \operatorname{tr} ( \chi^\dagger_{\ell} \chi^{\vphantom\dagger}_{\ell} )
+ m \sum_s \operatorname{tr} (\mu_s) + {\cal H}_F \text{,}
\end{equation}
where ${\cal H}_F = {\cal H}_K + {\cal H}_V$, and the latter two operators are constructed as in Eqs.~(\ref{eqn:hk},\ref{eqn:hv}).

We consider the following ansatz:
\begin{eqnarray} 
\chi^{a b}_{\ell} &=& - \delta^{a b} \chi \\
\mu^{a b}_s &=& - \delta^{a b} z_s \chi \text{.} \label{eq:kclusteransatz}
\end{eqnarray}
Here, $z_s$ is the coordination number of the site $s$.  We shall see that $\chi > 0$ upon minimizing the energy with respect to $\chi$.  With this choice, fixing the color and spin quantum numbers, the one-particle Hamiltonian that can be read off from ${\cal H}_F$ is proportional to the Laplacian matrix of the graph (with positive coefficient).  Therefore the single particle ground state (for fixed color and spin) has zero energy, is unique, and its wavefunction is a constant.  The unique many-particle ground state of ${\cal H}_F$ is obtained by filling this state with  $k m n_c = n_c N$ fermions, one in each of the  $n_c N$ possible combinations of color and spin states.  The mean-field energy is therefore given entirely by the constant terms in $H_{{\rm MFT}}$, and is
\begin{eqnarray}
E_{{\rm MFT}} &=& \frac{n_c N N_b}{{\cal J}} \chi^2 - m n_c \chi \sum_s z_s \\
&=&  \frac{n_c N N_b}{{\cal J}} \chi^2 - 2 m n_c N_b \chi \text{,}
\end{eqnarray}
where $N_b$ is the number of links in the graph.  Minimizing with respect to $\chi$, we find
\begin{equation}
\label{eqn:kcluster-energy}
E_{{\rm MFT}} = - \frac{n_c N N_b {\cal J}}{k^2} \text{.}
\end{equation}
We know this must be the ground state energy of the isolated $k$-cluster because it saturates the bound Eq.~(\ref{eqn:bond-gs-bound}) provided by the ground state energy of the two-site problem.  

We note that this result also holds at any finite $N$.  Schematically, this can be seen by noting that the ground state is the unique singlet that can be formed from the $k$ spins, which can be thought of as a $N \times n_c$ Young tableau, which is obtained by vertically stacking the $m \times n_c$ tableaux for each of the $k$ sites.  Any pair of spins can then be seen to transform in the $2m \times n_c$ representation, which, by the discussion of Appendix~\ref{app:2site-exact}, implies that the two-site Hamiltonian on the link connecting those sites is in its ground state.  So the ground state energy is just the sum of the two-site ground state energies for each link in the graph, which again gives Eq.~(\ref{eqn:kcluster-energy}).

\section{Chiral spin liquid in large-$k$ limit}
\label{app:largek}

Our demonstration that  constant magnetic field with a flux of $2\pi/k$ per plaquette is the lowest energy solution to the saddle point equations -- on the square lattice, for $n_c = 1$ and $5 \leq k \leq 8$ -- is purely numerical. A natural question which arises in this context is whether this can be supplemented by additional analytical analysis. 

The solution to the problem of a particle hopping on  a square lattice in a constant magnetic field of flux $2\pi/k$ cannot be found analytically. Yet it is well known that the spectrum consists of $k$ bands (Landau levels). \cite{hofstadter76}  Since the fermions we work with are at a filling fraction $1/k$ (filling all the bands would correspond to $N$ particles per site, while we have instead $m=N/k$ particles per site), they fill precisely one lowest Landau level. Yet the energy of a filled Landau level is not known analytically, except at a very large $k$ where the problem becomes effectively continuous. 

Therefore, let us calculate the energetics of a saddle point solution with a flux of $2\pi/k$ per plaquette (corresponding to ACSL) in the limit of very large $k$ and compare it with other possible states at this $k$.  Our analysis will be for the case $n_c = 1$, but also applies immediately to $n_c = 2$, since any $n_c = 1$ saddle point can be extended to a $n_c = 2$ saddle point of the diagonal form
\begin{eqnarray}
\chi^{a b}_{\br \br'} &=& \delta^{a b} \chi^a_{\br \br'} \text{   (no sum)} \\
\mu^{a b}_{\br} &=& \delta^{a b} \mu^a_{\br} \text{   (no sum),}
\end{eqnarray}
where each pair $(\chi^a_{\br \br'}, \mu^a_{\br})$ is a $n_c = 1$ saddle point solution.
The energy is simply a sum of energies of the $n_c = 1$ saddle points.  We can obtain the nACSL and dCSL saddle points in this fashion from the ACSL saddle point, by choosing $\chi^1$ and $\chi^2$ to have the same or opposite magnetic fields, respectively.  The other $n_c = 1$ states we consider can similarly be straightforwardly extended to $n_c = 2$ states in this fashion.

We start by choosing the hoppings $\chi_{\br \br'}$ according to Eq.~(\ref{eq:uniformfield}). 
First of all, the first term in the mean-field energy is easy to calculate
\begin{equation} \label{eq:bondssL} \frac{N}{\cal J} \sum_{\left< \br \br' \right>} \left| \chi_{\br \br' } \right|^2 = \frac{2 N_s N  \chi^2}{{\cal J} }.
\end{equation}
Here, as before, $N_s$ is the total number of sites on the lattice, and $2N_s$ is the total number of bonds. 

Now let us find the energy of a fully filled Landau level. 
A particle hopping on a lattice with a hopping strength $\chi$ without the magnetic field has the spectrum
\begin{eqnarray} \label{eq:dispersion}  \epsilon(k_x,k_y) & = &  -2 \chi \cos(k_x ) - 2\chi \cos(k_y ) \approx \cr && -4 \chi + \chi \left(k_x^2 +k_y^2 \right) - \frac \chi {12} \left(k_x^4+k_y^4 \right)
\end{eqnarray}
where a small $k_x$, $k_y$ expansion was performed (lattice spacing is taken to be unity). Looking at the quadratic term, we read off the effective mass of the particle $m^* = 1/(2 \chi)$. This gives the cyclotron frequency
\begin{equation} \omega = \frac{B}{m^*} = \frac{4\pi \chi}{k},
\end{equation}
since the magnetic field is $B=2\pi/k$. The energy of the lowest Landau level is then 
\begin{equation} \label{eq:landaulevel} E_L = -4 \chi + \frac 1 2 \omega = - 4 \chi + \frac{2 \pi \chi}{k}.
\end{equation}

For what follows we would like to also calculate the $1/k^2$ correction to this result. The corrections come from the quartic term in the dispersion, which takes into account the deviation of the lattice from the continuum limit. The correction to the Hamiltonian describing the motion of  a particle in a magnetic field in the continuum due to this term in the dispersion can be found by minimal subtraction (for example, in Landau gauge), and gives
\begin{equation} 
V = - \frac{\chi}{12}  \left[ \left( -i \frac{\partial}{\partial x} + \frac{2 \pi y}{k} \right)^4 +\frac{\partial^4}{\partial y^4} \right].
\end{equation}
Considering this a perturbation, the unperturbed wave function is given by \cite{LandauLifshitzQuantum}
\begin{equation} \psi(k_x) = \left( \frac 2 k  \right)^{\frac 1 4} e^{i k_x x} \exp \left( - \frac \pi k \left(y+ \frac{k k_x}{2 \pi} \right)^2 \right).
\end{equation}
Calculating the matrix element $\left< \psi(k_x) \right| V \left| \psi(k_x) \right>$ we find
\begin{equation} E_L = -4 \chi + \frac{2\pi \chi}k - \frac{\pi^2\chi}{2 k^2}.
\end{equation}
This is the energy of the lowest Landau level in the approximation up to terms $1/k^2$. Notice that the Landau level remains flat, that is, $k_x$ independent. It is easy to see that it will remain flat up to arbitrary order in $1/k$. This means that the broadening of the Landau level is exponentially small in $1/k$ and can be ignored for the purposes of this calculation.

The total number of particles filling the Landau level is $N N_s/k$, so the mean field energy becomes
\begin{equation}
E_{\rm MFT} = \frac{2 N_s N \chi^2}{\cal J} - \frac{N N_s E_L}{k}.
\end{equation}
Minimizing this with respect to $\chi$ we find
\begin{equation} \label{eq:magneticfields}
E_{\rm MFT} = - \frac{{\cal J} N N_s}{k^2} \left( 2 - \frac{2 \pi}{k} + \frac{ \pi^2}{k^2} + \dots \right).
\end{equation}

Now let us consider alternative states. One alternative state is a Fermi surface state, where all hoppings are real and equal to $\chi$. The energy of such a state is straightforward to calculate. We take particles moving with the dispersion given by Eq.~(\ref{eq:dispersion}), fill all the states at an appropriate density up to Fermi energy, to find the total energy per particle to be
\begin{equation}
E_{F} = -4 \chi + \frac{2\pi \chi}k - \frac{\pi^2\chi}{3 k^2}.
\end{equation}
This energy is slightly higher than the energy of the Landau level given in Eq.~(\ref{eq:landaulevel}). Therefore the energy after minimization
with respect to $\chi$ is also slightly higher
\begin{equation} \label{eq:fermisurface}
E_{\rm MFT} = - \frac{{\cal J} N N_s}{k^2} \left( 2 - \frac{2 \pi}{k} +  \frac{5  \pi^2}{6 k^2} + \dots \right).
\end{equation}
Clearly, Eq.~(\ref{eq:magneticfields}) is greater than Eq.~(\ref{eq:fermisurface}), so the state with the uniform magnetic field wins. 

A second alternative state one might consider is a VCS state. Suppose the lattice is covered by clusters of exactly $k$-sites each, each containing $N_b$ bonds.  Within each cluster
$\chi_{\br \br' } $ are constant and equal to $\chi$, which is real, and $\chi_{\br \br' }$ for bonds connecting the clusters are zero.  In Appendix~\ref{app:kcluster} it is found that the energy of a single cluster is given by Eq.~\ref{eqn:kcluster-energy}).  Since the cluster energies simply add, and the  number of clusters is $N_s / k$, the total energy is
\begin{equation}
E_{{\rm MFT}} = - \frac{{\cal J} N N_s N_b}{k^3} \text{.}
\end{equation}

Now we can define $N_{be}$ by
\begin{equation}
\frac{N_s N_{b}}{k} = 2 N_s - N_{be}  \text{,}
\end{equation}
so that  $N_{be}$ is the total number of bonds not contained inside some cluster.
We have
\begin{equation}
E_{\rm MFT} = - {\cal J} \frac{2 N N_s}{k^2} + {\cal J}  \frac{N N_{be}}{ k^2}.
\end{equation} 
$N_{be}$ scales with the total perimeter of all clusters.  Since the perimeter of a single large cluster goes like $\sqrt{k}$, $N_{be}$ is proportional to $\sqrt{k}$ times the total number of clusters $N_s / k$, or
\begin{equation} N_{be} = c \frac{N_s}{\sqrt{k}},
\end{equation}
where $c$ is some constant. This gives
\begin{equation}
E_{\rm MFT} = - \frac{ {\cal J} N N_s}{k^2} \left(2 - \frac{c}{ \sqrt{k}} \right).
\end{equation}
Comparing this with the uniform magnetic mean-field energy Eq.~(\ref{eq:magneticfields})
as well as the uniform hopping given by  Eq.~(\ref{eq:fermisurface}), we see that the magnetic field mean-field energy is again the lowest
at large $k$.

The arguments presented here do not prove that the uniform magnetic field is the lowest energy solution. That is demonstrated instead by the numerical solution of the mean-field equations. However, they do give a feel and perhaps some intuition as to why this solution wins over some of the possible alternatives. They also support the idea that chiral spin liquids are good ground states not just a few intermediate values of $k$, but also for larger $k$.  Therefore we conjecture that the ACSL is the large-$N$ ground state for $n_c = 1$ and all $k \geq 5$, and that the nACSL and dCSL are the degenerate large-$N$ ground states for $n_c = 2$ and all $k \geq 6$.

\section{Localization and braiding of fractional and non-Abelian particles}
\label{app:carriers}

One of the most striking properties of the topological liquid states discussed in this paper is the presence of particles with fractional and non-Abelian statistics.  It is therefore interesting to discuss how, in principle, such particles may be localized and braided, especially in view of the intense interest in topological quantum computation using non-Abelian particles.  Our intent here is not to develop a detailed and realistically achievable proposal to carry out such a braiding experiment in a cold atom system, but simply to discuss in principle how such braiding may be achieved, and point out some of the issues that arise.  Development of more detailed proposals is an interesting subject for future work.  It would be also interesting if our discussion can be sharpened by appropriate calculations.  For ease of presentation, we focus on the case $n_c = m  = 1$; generalization to other cases is straightforward.  

We shall discuss how one may localize a particle called a holon that is spinless but carries the conserved atom number.  The reason we consider holons rather than spinons, is that holons may be localized simply by modifying the optically generated single particle potential for the atoms.  To do this, we need to go beyond the Heisenberg spin model, and for greatest simplicity we consider a $t$-$J$ model where strong correlation restricts the number of atoms per site to be less than or equal to one.  The Hamiltonian is
\begin{eqnarray} H_{tJ} &=& - t_g \sum_{\left<\br  \br' \right>} {\cal P} \left( c^\dagger_{\br' \alpha } c_{\br \alpha} + {\rm H.c.} \right) {\cal P}  \nonumber \\ &+& J  \sum_{\left<\br \br' \right>}  c^\dagger_{\br' \alpha} c_{\br' \beta}
c^\dagger_{\br \beta} c_{\br \alpha} \text{,}
\end{eqnarray}
where $c^\dagger_{\br \alpha}$ creates a ground state atom in spin state $\alpha$ on site $\br$, and ${\cal P}$ is a projector onto the subspace with one or fewer atoms on each site.   $J_{\br \br'}$ has been replaced by $J$ on every bond, and the sum in the first term is over nearest-neighbor bonds of the square lattice. $c_{\br \alpha}$ is said to insert a hole (with spin $\alpha$) at site $\br$.  When there are no holes present, this model reduces to the Heisenberg spin model with $n_c = m = 1$.  Below, we rely on the approach of Lee and Nagaosa to discuss this model.\cite{nagaosa92} 

To make contact with the description of the topological liquid states, we decompose the hole insertion operator as
\begin{equation}
\label{eqn:tj-decomposition}
c_{\br \alpha} = f_{\br \alpha} b^\dagger_{\br} \text{,}
\end{equation}
where $f^\dagger_{\br \alpha}$ creates a spinon and $b^\dagger_{\br}$ is a bosonic creation operator creating a holon.  Spinon and holon densities obey the local constraint
\begin{equation}
\label{eqn:tj-constraint}
f^\dagger_{\br \alpha} f^{\vphantom\dagger}_{\br \alpha} + b^\dagger_{\br} b^{\vphantom\dagger}_{\br} = 1 \text{.}
\end{equation}

Assuming the system (without holes) has an ACSL ground state, the spinons are low-energy quasiparticles, which couple to the Chern-Simons gauge field and thus acquire fractional statistics.  The holon also carries gauge charge, and thus also acquires fractional statistics.  It should be noted that, in the equations above the holon and spinon are formal objects used to microscopically represent the $t$-$J$ model, and these formal objects are not the same as the low-energy quasiparticle degrees of freedom.  Holons and spinons emerge as low-energy degrees of freedom when we study the $t$-$J$ model starting from an appropriate mean-field theory, \cite{nagaosa92} and then including fluctuations.  Since the discussion here is only qualitative, and since the needed mean-field theory is very closely related to that introduced in Sec.~\ref{sec:models}, we shall not introduce it here.  It suffices to note that, at the mean-field level, both spinons and holons are free particles, which are minimally coupled to the fluctuating gauge field upon going beyond mean-field theory.

We now consider introducing the external potential
\begin{equation}
\delta H_{tJ} = - \sum_{\br} U(\br) c^\dagger_{\br \alpha} c^{\vphantom\dagger}_{\br \alpha}  \text{,}
\label{eqn:ext-pot}
\end{equation}
and adding a single hole into the system.  The sign in Eq.~(\ref{eqn:ext-pot}) is chosen so that a negative $U(\br)$ is an attractive potential for the added hole.  Up to an additive constant, we may use Eq.~(\ref{eqn:tj-constraint}) to re-express the potential as
\begin{equation}
\delta H_{tJ} =  \sum_{\br} U(\br) b^\dagger_{\br } b^{\vphantom\dagger}_{\br }  \text{.}
\end{equation}
We could have equally chosen the potential to couple to the spinons and not the holons; the above choice is  convenient, but is purely a convention.  For example, at the mean-field level, a change in the saddle point value of the Lagrange multiplier field enforcing the local constraint, will apportion the effect of $U(\br)$ between holons and spinons.  Therefore the system dynamically determines the effect of the physical external potential $U(\br)$ on holons and spinons.

When the hole is added, Eq.~\ref{eqn:tj-decomposition} tells us that we both add a holon and remove one spinon.  (The removed spinon should really be called a spinon hole, but for ease of discussion we will simply call it a spinon.)  We suppose that $U(\br)$ is negative, appreciable only in a small spatial region, and just strong enough to bind a particle.  Because $U(\br)$ couples to the conserved density, we expect it to bind a particle carrying atom number $-1$, but it is not obvious whether this particle will be a hole or a holon.  To understand this, the added holon and spinon will interact via some short-ranged potential.  This potential may be attractive or repulsive.  If the holon-spinon potential is attractive enough, the holon will be bound to the spinon, and they will be localized together by the external potential $U(\br)$.  In this case we have localized a hole, which is not a fractional particle.  On the other hand, if the holon-spinon potential is repulsive enough, a holon will be localized.  In the latter case, we can then manipulate the fractional holon by adiabatically changing the external potential $U(\br)$.  Multiple holons could be created by choosing $U(\br)$ to be a sum of several localized potentials.

Since the goal is to create and manipulate a fractional particle, what should be done if a hole is localized by the external potential?  One solution is to apply a time-varying Zeeman magnetic field, which will couple to the localized spinon and can be used to excite it to a delocalized state, leaving behind a localized holon.  If we do this to create a state with several localized holons, the delocalized spinon excitations will induce some errors when the holons are braided.  However, these spinon excitations can be made to relax by whatever cooling mechanism was used to prepare the state of several localized holes in the first place.  (Finding a cooling mechanism capable of achieving this for cold atom Mott insulators is a significant unsolved problem.  Solving it is a prerequisite for \emph{any} experiment probing fractional or non-Abelian statistics in such systems, which would have to be carried out at temperatures well below the bulk gap.)  While some spinons may relax back into the localized states and re-form holon-spinon bound states, because these states are localized, the rates for other relaxation processes (for instance, relaxation into low-energy edge excitations) are expected to dominate.

\bibliography{sun-long}

\end{document}